\documentclass[prl,aps,amssymb,twocolumn,nofootinbib,superscriptaddress]{revtex4-1}
\usepackage{amsmath}
\usepackage{amssymb}
\usepackage{amsthm}
\usepackage{amsfonts}
\usepackage{listings}
\usepackage{enumerate}
\usepackage{latexsym}

\usepackage{psfrag}
\usepackage{hyperref}

\usepackage{bm}
\usepackage{graphicx}
\usepackage[caption=false]{subfig}
\usepackage{color}

\usepackage{multibib}

\newcommand{\beq}{\begin{equation}}
\newcommand{\eneq}{\end{equation}}
\input{epsf}

\newcommand{\supp}{\cite{SuppMat}}

\begin{document}

\tolerance 10000

\newcommand{\vk}{{\bf k}}

%\draft

\title{The chiral anomaly factory: Creating Weyl fermions with a magnetic field}

\author{Jennifer Cano}
\thanks{These authors contributed equally to the preparation of this work.}
\affiliation{Princeton Center for Theoretical Science, Princeton University, Princeton, NJ 08544}
\author{Barry Bradlyn}
\thanks{These authors contributed equally to the preparation of this work.}
\affiliation{Princeton Center for Theoretical Science, Princeton University, Princeton, NJ 08544}
\author{Zhijun Wang}
\affiliation{Department of Physics, Princeton University, Princeton, NJ 08544}
\author{Max Hirschberger}
\affiliation{Department of Physics, Princeton University, Princeton, NJ 08544}
\author{N. P. Ong}
\affiliation{Department of Physics, Princeton University, Princeton, NJ 08544}
\author{B. A. Bernevig}
\affiliation{Department of Physics, Princeton University, Princeton, NJ 08544}

% \author{Barry, Jen, Zhijun, Andrei and others?}

\date{\today}

\begin{abstract}
Weyl fermions can be created in materials with both time reversal and inversion symmetry by applying a magnetic field, as evidenced by recent measurements of anomalous negative magnetoresistance.
Here, we do a thorough analysis of the Weyl points in these materials: by enforcing crystal symmetries, we classify the location and monopole charges of Weyl points created by fields aligned with high-symmetry axes. 
The analysis applies generally to materials with band inversion in the $T_d$, $D_{4h}$ and $D_{6h}$ point groups.
For the $T_d$ point group, we find that Weyl nodes persist for \emph{all} directions of the magnetic field.
Further, we compute the anomalous magnetoresistance of field-created Weyl fermions in the semiclassical regime. We find that the magnetoresistance can scale non-quadratically with magnetic field, in contrast to materials with intrinsic Weyl nodes. 
Our results are relevant to future experiments in the semi-classical regime.

\end{abstract}

\maketitle

\section{Introduction} 
Following their insulating counterparts\cite{KaneReview}, topological semi-metals have attracted much recent theoretical and experimental interest. Weyl and Dirac semimetals have recently been theoretically predicted\cite{WangSunChenEtAl2012,WangWengWuEtAl2013,LiuJiangZhouEtAl2014,LiuZhouZhangEtAl2014,WengFangFangEtAl2015} and experimentally observed\cite{Ali14,YangLiuSunEtAl2015,LvXuWengEtAl2015}. Both display topologically protected Fermi-arc surface states, as well as large negative magnetoresistance due to the ``chiral anomaly''\cite{WanTurnerVishwanathEtAl2011,Burkov2011,Burkov15,Xiong15,Huang15,Zhang16}. While Weyl fermions in these semimetals are robust to small perturbations due to their topological character, Dirac points require a combination of crystal symmetry, time reversal, and inversion symmetry\cite{Yang14}. 
This suggests that Weyl fermions can be engineered by breaking inversion or time reversal symmetry in materials with four-band crossings.
While breaking inversion symmetry can be accomplished by adding strain\cite{Ruan15}, it is more straight-forward to break time reversal symmetry by turning on a magnetic field.
This route to creating Weyl fermions has already been carried out in GdPtBi\cite{Hirschberger16,Felser2016}, NdPtBi\cite{Felser2016}, Na$_3$Bi\cite{Xiong15} and Cd$_3$As$_2$\cite{Liang15,Li15,Li16}. We predict that the same route can be used to observe Weyl fermions in the experimentally relevant materials HgTe\cite{Bernevig06,Koenig07} and InSb\cite{Mourik12}.

Here, we consider turning on a magnetic field in materials with four-band crossings. We consider two types of four-band crossings: symmetry-enforced band crossings at the $\Gamma$ point and Dirac points near the $\Gamma$ point on a high-symmetry line. In both cases, the magnetic field breaks the four-band crossing into an even number of Weyl nodes.
We demonstrate the emergence of Weyl nodes explicitly in GdPtBi, HgTe and InSb, which host a symmetry-enforced four-band crossing near the Fermi level, and Cd$_3$As$_2$ and Na$_3$Bi, which host Dirac points near the Fermi level, using $k\cdot p$ Hamiltonians generated by ab initio calculations. We then show that this is a general result when the magnetic field is along a high-symmetry line.
However, the emergent Weyls do not always reside on the axis parallel to the magnetic field: instead, a complex map of Weyl points (and nodal lines\cite{Burkov11,ChiuSchnyder2014,ChanChiuChouEtAl2015,Fang15}) emerges for different directions of the magnetic field. Since Weyl points do not require symmetry protection, they persist when the magnetic field is moved away from these axes. Surprisingly, we find that in GdPtBi and HgTe, Weyl points exist for all directions of the magnetic field.
In the Supplement, we give numerical evidence for this statement and then prove it explicitly for all materials with $T_d$ symmetry and $j=3/2$ orbitals at the Fermi level.

We emphasize that while we use the language of applying an external magnetic field, our results apply equally well to magnetically ordered materials that host a four-band crossing above the N{\'e}el temperature and can thus be tuned to display Weyl points below this temperature. This is relevant for antiferromagnetic Heuslers\cite{Felser2016}.

Finally we consider the semiclassical negative magnetoconductance that is a consequence of the chiral anomaly. This has previously been considered for Weyl points that exist independent of the magnetic field\cite{sonspivak,spivak2}. Here, we show that when the Weyl points are created by a magnetic field, the magnetoconductance takes a different scaling form. In particular, it can scale as high as $B^{5/2}$, where the exact scaling depends on the linearized Hamiltonian near the Weyl point. We apply this model to GdPtBi, in which recent experiments\cite{Hirschberger16} have observed the chiral anomaly.

\section{Emergent Weyl nodes from a Symmetry-enforced four-band crossing}

Here we focus on GdPtBi, which is in the $T_d$ point group. However, because our analysis depends only on symmetry and band inversion, it applies to all materials with the same symmetry and relevant orbitals near the $\Gamma$ point, e.g., HgTe and InSb.

In GdPtBi, the low-energy spectrum near the $\Gamma$ point is described by the four $p$-orbitals with $j_z=\pm 3/2, \pm 1/2$. Thus, the symmetry operators comprise a four-dimensional representation of the $T_d$ symmetry group.
The $k\cdot p$ Hamiltonian takes the form\supp,
\begin{eqnarray}
&H_{\Gamma}(k_x, k_y, k_z)= \left( A_0+ (A+ \frac{C}{\sqrt{3}}) k^2 \right) I_{4 \times 4}  + \nonumber \\ 
 &+ C(k_x^2- k_y^2) \Gamma_1 + \frac{C}{\sqrt{3}} (2 k_z^2 - k_x^2- k_y^2) \Gamma_2 + \nonumber \\ 
& + E(k_x k_y \Gamma_3 + k_x k_z \Gamma_4 + k_y k_z \Gamma_5) + \nonumber  \\
& + D (k_x U_1 + k_y U_2 + 2 k_z U_3),  \label{kdotpzerofield}
\end{eqnarray}
where the $\Gamma_{1,...,5}$ are Clifford algebra matrices, $\Gamma_{ij} = [\Gamma_i,\Gamma_j]/(2i)$ and $U_1=\sqrt{3}\Gamma_{15}-\Gamma_{25}, U_2=-\sqrt{3}\Gamma_{14}-\Gamma_{24},U_3=\Gamma_{23}$. The parameters $A_0,A,C,D,E$ are obtained by an ab initio fit and given in the Supplementary Material. We note here only that $A_0\approx 0$, meaning that at the Fermi level
the spectrum is four-fold degenerate at the $\Gamma$ point\cite{Hirschberger16}; two bands disperse upwards and two disperse downwards. The $D\neq 0$ parameter breaks inversion symmetry.

We now consider what happens in the presence of a magnetic field by adding an effective Zeeman coupling to Eq~(\ref{kdotpzerofield}),
\begin{equation} H_Z = \vec{B}\cdot\vec{J},
\end{equation}
where $\vec{J}$ is a vector of the spin-$3/2$ matrices.
Using this model, oppositely-dispersing bands have either a protected or avoided crossing, \emph{regardless} of their $g$-factor. Notice that band inversion is crucial here: if all bands dispersed in the same direction, then the presence or absence of a crossing would depend crucially on the precise values $g$-factors of the bands. However, as long as the system exhibits band inversion, and all bands have the same sign of the $g$-factor, then the physics described below is universal. In the following analysis, we will ignore the effects of cyclotron motion to lowest order, focusing purely on the Zeeman splitting. This is a good approximation for large $g-$factor materials such as GdPtBi.

When the magnetic field is along an axis of rotation, band crossings along this axis between bands with different eigenvalues under the rotation are protected; thus, Weyl points are guaranteed to exist. Because the original four-band quadratic crossing was at the Fermi energy at zero field, these Weyl points will also lie near the Fermi level.
Additionally, since the nontrivial Chern number of a Weyl point cannot disappear under small deformations of the Hamiltonian, these Weyls will continue to exist even as the magnetic field is moved away from the rotation axis. This protection is crucial to experimental observation. 
However, as the magnetic field is moved far away from a high-symmetry axis, it is possible for two Weyl points of opposite chirality to meet and annihilate.
Surprisingly, we have verified numerically, using the $k\cdot p$ Hamiltonian (\ref{kdotpzerofield}), that this is not the case for every pair of nodes: Weyl points exist in GdPtBi for all directions of the magnetic field.

Weyl points can also be protected between high-symmetry planes with different Chern numbers; we show\supp  that in GdPtBi, when the magnetic field is applied along the $\hat{z}$ direction, two such Weyl points exist in the $k_x=0$ plane and another pair in the $k_y=0$ plane. These points persist --albeit moving in momentum-space-- when the magnetic field is moved off this axis until they reach each other and annihilate. 
The same analysis applies to a magnetic field in the $\hat{x}$ and $\hat{y}$ directions.

When the magnetic field is in the $\hat{x}+\hat{y}$ direction, we find four Weyl nodes which, for $D\rightarrow 0$, are confined to the $(k,k,k_z)$ plane; these nodes persist as $D$ is continuously varied from zero to its experimentally relevant value in either GdPtBi or HgTe. Additionally, at least one line node appears in the $(k,-k,k_z)$ plane.\supp For the ab initio parameters, two line nodes appear; however, for other choices of parameters, one of these lines moves towards the edge of the Brillouin zone and disappears.

Last, when the magnetic field is in the $\hat{x}+\hat{y}+\hat{z}$ direction, there are four Weyl points along the $k_x=k_y=k_z$ axis.

%\textbf{To check that these crossings are not an artifact of the $k\cdot p$ model, we verified their existence directly using a tight-binding model.}

Additional Weyl points at generic points in the Brillouin zone must occur in multiples of six or eight, depending on the symmetry that remains when a particular magnetic field is present. A summary of Weyl points that emerge in GdPtBi upon applying a magnetic field along a high-symmetry axis are shown in Table~\ref{tab:GdPtBi}. 

\begin{table}
\begin{center}
\begin{tabular}{ c | l }
  \hline			
$\vec{B}$ & Emergent Weyl points\\
\hline
$[001],[010],[100]$& $6 + 8k_1$ Weyl points \\
\hline
\begin{tabular}{c} $[110],[1\bar{1}0],[101],$\\$[10\bar{1}],[011],[01\bar{1}]$\end{tabular}& \begin{tabular}{l}0, 1, or 2 line nodes,\\
$4+8k_2$ Weyl points \end{tabular}\\
\hline
\begin{tabular}{c}$[111]$,\\ $[11\bar{1}],[1\bar{1}1],[\bar{1}11],$\\ $[\bar{1}\bar{1}1],[\bar{1}\bar{1}1],[1\bar{1}\bar{1}]$\end{tabular}& $4+6k_3$ Weyl points\\
  \hline  
\end{tabular}
\end{center}
\caption{Weyl points in GdPtBi when a magnetic field is applied along one of the high-symmetry axes in the first column. The integers $k_i$ indicate Weyl points that appear at generic points Brillouin zone; for GdPtBi, a numerical analysis of our $k\cdot p$ model yields $k_1=k_2=0, k_3=1$, but these numbers are material-dependent. }\label{tab:GdPtBi}
\end{table}

The symmetry-protected Weyl points we describe in this section also persist for arbitrary magnetic fields. In particular, at the special point $E=2C,D=0$ in Eq~(\ref{kdotpzerofield}), the Hamiltonian $H_\Gamma+H_Z$ is exactly solvable and has Weyl points along the momentum axis parallel to the magnetic field. We show in the Supplement that as $E$ is moved away from this fine-tuned value, the Weyl points move in space, but do not annihilate; we confirm this claim with a numerical analysis. Because the Weyl points are topologically protected, they will also persist for small values of $D$.

\section{Emergent Weyl nodes from Dirac points near the $\Gamma$ point}

In Cd$_3$As$_2$ and Na$_3$Bi, a different, but similar, scenario develops, where again band inversion plays a crucial role. There are two pairs of relevant orbitals at the $\Gamma$ point near the Fermi level: the $s$-orbitals with $j_z=\pm 1/2$ and the $p$-orbitals with $j_z=\pm 3/2$; each pair transforms as a two-dimensional representation of $D_{nh}$, where $n=4$ in Cd$_3$As$_2$ and $n=6$ in Na$_3$Bi. The two representations have different energies at the $\Gamma$ point, but, because the bands are inverted, they can cross elsewhere in momentum space. In the materials of interest, the crossings occurs at the Fermi level and are protected by $C_{nz}$ symmetry. Furthermore, the crossings occur close enough to the $\Gamma$ point to be described by the effective $k\cdot p$ model,
\begin{eqnarray}
&H_{\rm Dirac}(k_x,k_y,k_z)= \left(C_0+C_1k_z^2+C_2k_\parallel^2\right)I_{4\times 4}+\label{eq:kdotpCdAs}\\
&\tau_z\otimes \left( \left(M_0 + M_1k_z^2 + M_2k_\parallel^2\right)\sigma_z + A(\sigma_x k_x + \sigma_y k_y)\right),\nonumber
%&+ \left(\sigma_z \otimes\sigma_z\right)\left(M_0 + M_1k_z^2 + M_2k_\parallel^2\right) + \left(\sigma_z\otimes \sigma_x\right)Ak_x + \left(\sigma_z\otimes \sigma_y\right)Ak_y 
\end{eqnarray}
where $k_\parallel^2 = k_x^2+k_y^2$. 
Eq~(\ref{eq:kdotpCdAs}) describes two Dirac points on the $k_z$ axis at $k_z = \pm\sqrt{-M_0/M_1}$ ($M_0$ and $M_1$ have opposite signs in the ab initio fit; the fitting parameters for the quadratic terms and the symmetry-allowed third order terms are included in the Supplementary Material.) Because the four relevant bands come in pairs with distinct angular momentum character, we allow for two different g-factors $g_s$ and $g_p$ in the Zeeman coupling\supp.
In the presence of a magnetic field, each Dirac point can split into up to four Weyl points. As in the previous section, putative crossings will exist regardless of the $g$-factors of the different orbitals because the bands disperse in opposite directions. Furthermore, Weyl points that emerge when the magnetic field is along a particular high-symmetry axis must persist when 
the field is slightly off-axis because they can only annihilate in pairs of opposite chirality.

We now summarize our results\supp.
When the magnetic field is along the $\hat{z}$ direction all band crossings are protected by $C_{nz}$ symmetry: depending on the value of the magnetic field, this implies between four and eight Weyl points.
Band crossings between the $j_z=\pm 3/2$ and $\mp 1/2$ bands are double Weyl points -- these are not robust to small changes in the magnetic field; instead, they can split into two single Weyl points.
Line nodes, protected by $M_{001}$, can also emerge in the $k_z=0$ plane for large enough magnetic field.

When the magnetic field is along the $\hat{x}$ direction, $C_{2x}$ symmetry can protect between two and four total Weyl points and $M_{100}$ symmetry protects line nodes in the $k_x=0$ plane.  The same is true for the other symmetry-related directions.

\section{Semiclassical Magnetotransport}
The presence of Weyl points near the Fermi surface in a material - whether intrinsic or created by an external field - leads to an experimentally measurable negative magnetoresistance.\cite{Xiong15,Huang15,Zhang16,Hirschberger16}
The origin of this effect is due to the non-trivial Berry curvature surrounding each Weyl node, and is a manifestation of the so-called ``chiral anomaly.'' Previous theoretical analysis of this effect has been carried out for materials with intrinsic, field independent Weyl nodes, in both the semiclassical and ultra-quantum (only the lowest Landau level occupied) limits.\cite{sonspivak,spivak2,Burkov15,NN} For these intrinsic Weyls, chiral kinetic theory implies that, in the semiclassical limit and at low temperature, there is an anomalous positive magnetoconductance of the form
\begin{equation}
\sigma_a^{\mu\nu}=\sum_i\frac{\tau v_i^3}{8\pi^2\mu_i^2}B^\mu B^\nu, \nonumber
\end{equation}
where $\tau$ is the inter-nodal scattering rate, $v_i$ is the (geometric) mean velocity of node $i$, and $\mu_i$ is the chemical potential measured from node $i$. We wish to generalize this result to the case where the Weyl nodes are created by an external magnetic field. Relegating the details of the derivation to the Supplementary Material, we find that the magnetoconductance acquires additional field dependence due to the field dependence of the Weyl velocities, which enters both explicitly and through the now-strongly-field-dependent scattering time. We find for low fields
\begin{equation}
\sigma^{\mu\nu}_{a}=\sum_i\frac{\tau^i(B)|\det A^i|}{8\pi^2\mu_i^2}B^\mu B^\nu (1+\mathcal{O}(B)). \label{eq:conductance}
\end{equation}
where $A^i$, along with the vector $u^i$, which, in the materials we consider, enters the $\mathcal{O}(B)$ corrections, parameterize the linearized two-band Hamiltonian near the Weyl point at $k^i$:\cite{Soluyanov15}
\begin{equation} H  =u_j^i (k_j-k^i_j) \mathbb{I} + (k_j-k^i_j)A^i_{jk} \sigma_k \label{eq:twoband}
\end{equation}
%up to a constant,
%where $u_i$ and $A_{ij}$ parameterize the Weyl points and the $\sigma_j$ are the Pauli matrices. 
$\tau^i(B)$ is the rate for scattering out of node $i$. For short-range impurities we find 
\begin{equation}
\tau^i(B)=\tau_0\frac{2\pi^2|\det A^i|}{\mu_i^2}(1+\mathcal{O}(B))
\end{equation}

We now make three observations. First, because $|\det A^i|$ depends on magnetic field, we expect that the magnetoconductance for field-created Weyls scales differently than that for intrinsic Weyl semimetals; in particular, we provide a simple model in the Supplementary Material where $|{\rm det} A^i | \sim |B|^{1/2}$. Second, because $|\det A^i|$ is not a rotationally invariant function of $\mathbf{B},$ we expect that the magnetoconductance will also fail to be rotationally invariant. Finally, as $\mathbf{B}$ increases, we expect the $\mathcal{O}(B)$ corrections to Eq. (\ref{eq:conductance}), given explicitly in the Supplementary material, to become significant. This can cause the directionality of the magnetoconductance to acquire additional field-dependence.

Similarly, thermoelectric transport in Weyl materials is also influenced by the chiral anomaly. In particular, the thermoelectric conductivity $\alpha^{\mu\nu}$, relating the current response to a temperature gradient, is experimentally relevant. Using Onsager reciprocity\cite{Onsager1931,Onsager1931-1}, we can compute this by looking instead at the energy current response to an electric field. Using the same semiclassical treatement as above and neglecting interactions we recover the Mott formula
\begin{equation}
\alpha^{\mu\nu}=\frac{\pi^2T^2}{3}\frac{\partial\sigma^{\mu\nu}}{\partial \mu},
\end{equation}
valid for both the anomalous and non-anomalous parts of the thermoelectric conductivity. In particular, we expect that the magnetic field dependence of the anomalous thermoelectric conductivity should simply follow that of the ordinary conductivity. Differences between these two effects serves to measure the significance of electron-electron interactions, which explicitly modify the heat-current\cite{Jonson1980}.

Lastly, we remark on the effects of higher-order Weyl crossings on magnetotransport. In particular, we focus on double-Weyl points, since -- as mentioned above -- these are present in $\mathrm{Na_3Bi}$ and $\mathrm{Cd_3As_2}$. Using the fact that the Berry curvature transforms as a tensor under reparametrizations of the Brillouin zone, we can easily repeat the semiclassical analysis above for double (or even $n$-fold) Weyl points. We find that the forms of all transport coefficients remain the same. The only change is that the response coefficients are proportional to the square of the Chern number (i.e. $4$ in the case of a double Weyl), and that the form of the density of states changes. In particular, the density of states for a double Weyl point is linear in the chemical potential.

\section{Validity of semiclassical transport}

We now consider whether Weyl points can be well separated when the system is in the semiclassical regime. This introduces two competing criteria. First, the Weyl points must be well-resolved: the Fermi level must be close enough to the nodal point that the Fermi surface consists of disconnected pockets encircling each node. Quantitatively, this translates to the constraint,
\begin{equation}
k_F\sim \frac{\mu}{v}\ll k^0
\label{eq:separated}
\end{equation}
where $v=(\det A)^{1/3}$ is the mean velocity of the Weyl point at position $k^0$, $k_F$ is the Fermi wavevector measured as the deviation from $k^0$, and the chemical potential $\mu$ is the deviation in energy from the Weyl point. 

Second, we demand that the number $\nu$ of filled Landau levels is large. Recall that for a single Weyl point,
\begin{equation}
\mu\sim\sqrt{2B\nu}.
\end{equation}
Hence, we demand,
\begin{equation}
\mu \gg \sqrt{2B}
\label{eq:manyLL}
\end{equation}

We now consider when the two constraints~(\ref{eq:separated}) and (\ref{eq:manyLL}) are simultaneously satisfiable. 
%Whether or not these two constraints are simultaneously satisfiable depends upon the scaling relation between $v,k^0$ and $B$, as well as the material parameters. 
For Weyl points that originate from a symmetry-enforced band touching, such as those in GdPtBi, $v\sim k^0\sim \sqrt{B}$, and hence we need simultaneously that
\begin{equation}
B\gg \gamma_1 \mu \text{ and }B\ll \gamma_2\mu^2 \label{eq:semiconstraints}
\end{equation}
where $\gamma_1$ and $\gamma_2$ are material-dependent parameters. 
Whether or not there exists a regime that satisfies Eq~(\ref{eq:semiconstraints}) depends on $\mu$, which is nearly fixed (for $B$ not too large) for a bulk $3d$ material.
For GdPtBi, we find from experiment\cite{Hirschberger16} that the Weyl points become well resolved for $B\sim 6T$; however quantum oscillations reveal that the Landau level index $\nu\sim 5$ at this value of the field.
However, the preceding analysis ignores the magnetization of GdPtBi. Near the N\'{e}el temperature of about 9K, the spins will have a large magnetic susceptibility, in which case a smaller field will have the same effect. Additionally, this will be compounded by the quenching of the orbital magnetism in the crystal, leading to an enhancement of the Zeeman energy relative to the cyclotron energy. In this case, it is quite possible that the experimental regime satisfies Eq~(\ref{eq:semiconstraints})\cite{Felser2016}.

If the scale of inversion breaking, $D$ in Eq~(\ref{kdotpzerofield}), is much larger than the magnetic field, then $k^0\sim\sqrt{B}$ and $v\sim D$. Then Eq~(\ref{eq:semiconstraints}) is replaced by,
\begin{equation} \gamma_1\mu \ll D\sqrt{B} \text{ and } \sqrt{B} \ll \gamma_2\mu,\end{equation}
which is satisfied for small enough fields when $D$ exceeds other scales.

We now consider Weyl points that emerge from splitting a Dirac point with a magnetic field.
In this case, for the two Weyl points to be well-resolved, the spacing between the Weyl points, which scales like $B$, must be greater than $k_F \sim \mu/v$. To leading order, $v$ depends on the initial dispersion (i.e., determined by Eq~(\ref{eq:kdotpCdAs})) and only has sub-leading $B$ dependence. This again leads to Eq~(\ref{eq:semiconstraints}). 
The recent experiment on Na$_3$Bi\cite{Xiong15} reports the nodes to be well separated when $B=12T$, but the onset of the lowest Landau level to be near $6-8T$. Hence, the semiclassical regime will likely not quantitatively describe this experiment.

\section{Discussion}

A magnetic field can create Weyl points from four-band crossings by breaking time reversal symmetry.
This is a powerful technique for creating Weyl points whose position in energy-momentum space is tunable. 
Here, we have studied two canonical and experimentally relevant examples: a symmetry-enforced four-band crossing at the $\Gamma$ point and a Dirac node near the $\Gamma$ point.
We have shown that a complex map of Weyl points (and line nodes) emerges, depending on the direction and magnitude of the magnetic field. It would be interesting to experimentally track the movement of these points by observing how the surface Fermi arcs move as the magnetic field is changed, e.g., in STM experiments. Furthermore, for the particular cases of GdPtBi and HgTe, our numerical analysis indicates that Weyl points exist near the Fermi level for all directions of the magnetic field: this should prompt future experiments that probe the chiral anomaly with fields away from the high-symmetry axes.

We computed the anomalous longitudinal conductance in the semiclassical regime for Weyl points created by a magnetic field. The conductance scales with a higher power of the magnetic field than the conductance for intrinsic Weyl points. 
Naively, this is consistent with experimental data: for example, the low-field data in Ref~\onlinecite{Hirschberger16} shows that $\sigma_{xx}$ scales like a higher power of $B$ than $B^2$; we plot this data in the Supplement.
However, this agreement should be taken with a grain of salt, because, as mentioned in the previous section, the experiment is not fully in the semi-classical regime.
%This is consistent with experimental observation: for example, for intrinsic Weyl points, when the magnetic field is rotated with respect to the electric field by some angle, $\theta$, the longitudinal conductance is proportional to $\cos^2\theta$, while Ref~\onlinecite{Xiong15} reports a fit to $\cos^4\theta$ and, similarly, Fig 2b in Ref~\onlinecite{Hirschberger16} is better fit by $\cos^3\theta$ or $\cos^4\theta$ than $\cos^2\theta$. However, there are two details swept under the rug in this naive analysis: first, as mentioned in the previous section, the experiments are not fully deep in a clean semiclassical regime; and second, we have demonstrated that the Weyl points move as a function of magnetic field and thus would not expect the data to be described by any power of $\cos\theta$.
Our theory will be better tested in future experiments that are in this regime, where we expect the scaling of longitudinal conductance to go beyond $B^2$ for magnetic-field created Weyl points.
%Our main point is that the scaling of longitudinal conductance like $B^2$ for intrinsic Weyl points does not apply to magnetic-field created Weyl points and we expect this to be crucial in understanding future experiments.

Our analysis readily generalizes to other point groups. This would be a useful course of study to identify future candidates for magnetic field created Weyl points. In addition, an analysis of the quantum regime could be used to describe existing experiments.  We leave these questions for future works.

\section{Acknowledgements}
The authors thank Alexey Soluyanov and Claudia Felser for helpful discussions. 
BAB acknowledges the hospitality and support of the Donostia International Physics Center and the \'{E}cole Normale Sup\'{e}rieure and Laboratoire de Physique Th\'{e}orique et Hautes Energies and the support of the Department of Energy de-sc0016239, NSF EAGER Award  DMR -- 1643312, Simons Investigator Award, ONR-N00014-14-1-0330, ARO MURI W911NF12-1-0461, NSF-MRSEC DMR-1420541, the Packard Foundation, the Keck Foundation, and the Schmidt Fund for Innovative Research.

\bibliography{kdotp}

%%%%%%%%%% Merge with supplemental materials %%%%%%%%%%
\clearpage
\begin{widetext}
\begin{center}
\textbf{\large Supplementary Material for The chiral anomaly factory: Creating Weyl fermions with a magnetic field}
\end{center}
\end{widetext}
%%%%%%%%%% Merge with supplemental materials %%%%%%%%%%
%%%%%%%%%% Prefix a "S" to all equations, figures, tables and reset the counter %%%%%%%%%%
\setcounter{equation}{0}
\setcounter{figure}{0}
\setcounter{table}{0}
\makeatletter
\renewcommand{\theequation}{S\arabic{equation}}
\renewcommand{\thefigure}{S\arabic{figure}}
\section{4D Irreps at $\Gamma$: G\lowercase{d}P\lowercase{t}B\lowercase{i}, H\lowercase{g}T\lowercase{e}, I\lowercase{n}S\lowercase{b}}
\label{sec:GdPtBi}
In this Appendix, we perform the detailed symmetry analysis of materials with a quadratic four-band crossing at the $\Gamma$ point in a Zeeman field. We will first derive the most general low-energy $\mathbf{k}\cdot\mathbf{p}$ Hamiltonian near the $\Gamma$ point. We then deduce the location and multiplicities of all sets of Weyl points that emerge when the magnetic field is aligned with high-symmetry crystallographic directions.

%The low-energy bands in GdPtBi come from the Bi $6p$ orbitals\cite{Hirschberger16}. 
In all cases we consider, the low-energy bands come from the $p$ orbitals with total angular momentum $J=3/2$.
We have chosen the ordered basis $|J,m_j\rangle = |\frac{3}{2},\frac{3}{2}\rangle,|\frac{3}{2},\frac{1}{2}\rangle,|\frac{3}{2},-\frac{1}{2}\rangle,|\frac{3}{2},-\frac{3}{2}\rangle$.
The $J={3/2}$ angular momentum operators are:
\beq
J_x= \left(
\begin{array}{cccc}
 0 & \frac{\sqrt{3}}{2} & 0 & 0 \\
 \frac{\sqrt{3}}{2} & 0 & 1 & 0 \\
 0 & 1 & 0 & \frac{\sqrt{3}}{2} \\
 0 & 0 & \frac{\sqrt{3}}{2} & 0 \\
\end{array}
\right) \label{Jx}
\eneq
\beq
J_y = \left(
\begin{array}{cccc}
 0 & -\frac{1}{2} \left(i \sqrt{3}\right) & 0 & 0 \\
 \frac{i \sqrt{3}}{2} & 0 & -i & 0 \\
 0 & i & 0 & -\frac{1}{2} \left(i \sqrt{3}\right) \\
 0 & 0 & \frac{i \sqrt{3}}{2} & 0 \\
\end{array}
\right)\label{Jy}
\eneq
\beq J_z=\left(
\begin{array}{cccc}
 \frac{3}{2} & 0 & 0 & 0 \\
 0 & \frac{1}{2} & 0 & 0 \\
 0 & 0 & -\frac{1}{2} & 0 \\
 0 & 0 & 0 & -\frac{3}{2} \\
\end{array}
\right)\label{Jz}\eneq

\subsection{Symmetries}
In this basis, we can represent the elements of the symmetry group $T_d$ along with time reversal by the matrices ($K$ represents complex conjugation)

\beq
T= \left(
\begin{array}{cccc}
 0 & 0 & 0 & -1 \\
 0 & 0 & 1 & 0 \\
 0 & -1 & 0 & 0 \\
 1 & 0 & 0 & 0 \\
\end{array}
\right) K
\eneq

\beq
C_{2x}= \left(
\begin{array}{cccc}
 0 & 0 & 0 & i \\
 0 & 0 & i & 0 \\
 0 & i & 0 & 0 \\
 i & 0 & 0 & 0 \\
\end{array}
\right)
\eneq

\beq
C_{2y} =\left(
\begin{array}{cccc}
 0 & 0 & 0 & -1 \\
 0 & 0 & 1 & 0 \\
 0 & -1 & 0 & 0 \\
 1 & 0 & 0 & 0 \\
\end{array}
\right)
\eneq

\beq
C_{2z}= \left(
\begin{array}{cccc}
 i & 0 & 0 & 0 \\
 0 & -i & 0 & 0 \\
 0 & 0 & i & 0 \\
 0 & 0 & 0 & -i \\
\end{array}
\right) \label{c2z}
\eneq
\beq
C_{4z}I = \left(
\begin{array}{cccc}
 -\sqrt[4]{-1} & 0 & 0 & 0 \\
 0 & -(-1)^{3/4} & 0 & 0 \\
 0 & 0 & \sqrt[4]{-1} & 0 \\
 0 & 0 & 0 & (-1)^{3/4} \\
\end{array}
\right)
\eneq
\beq 
M_{110}= \left(
\begin{array}{cccc}
 0 & 0 & 0 & -(-1)^{3/4} \\
 0 & 0 & \sqrt[4]{-1} & 0 \\
 0 & (-1)^{3/4} & 0 & 0 \\
 -\sqrt[4]{-1} & 0 & 0 & 0 \\
\end{array}
\right)
\eneq
\beq
C_{3, 111}= 
\frac{\sqrt{2}}{4}e^{-i\pi/4}
\left(\begin{array}{cccc}
-i & -\sqrt{3} & i\sqrt{3} & 1\\
-i\sqrt{3} & -1 & -i & -\sqrt{3}\\
-i\sqrt{3} & 1 & -i & \sqrt{3}\\
-i & \sqrt{3} & i\sqrt{3} & -1
\end{array}\right)
\eneq
%\beq
%\left(
%\begin{array}{cccc}
 %-\frac{1+i}{4} & \left(-\frac{1-i}{4}\right) \sqrt{3} & \left(\frac{1+i}{4}\right) \sqrt{3} & \frac{1-i}{4} \\
 %\left(-\frac{1+i}{4}\right) \sqrt{3} & -\frac{1-i}{4} & -\frac{1+i}{4} & \left(-\frac{1-i}{4}\right) \sqrt{3} \\
 %\left(-\frac{1+i}{4}\right) \sqrt{3} & \frac{1-i}{4} & -\frac{1+i}{4} & \left(\frac{1-i}{4}\right) \sqrt{3} \\
 %-\frac{1+i}{4} & \left(\frac{1-i}{4}\right) \sqrt{3} & \left(\frac{1+i}{4}\right) \sqrt{3} & -\frac{1-i}{4} \\
%\end{array}
%\right)
%\eneq
Notice the unconventional $C_{4z}I$ symmetry, where $I$ is inversion; inversion by itself is not a symmetry.
The Hamiltonian $H(k_x, k_y, k_z)$ then must satisfy:
\beq
T H(k_x, k_y, k_z) T^{-1} = H(-k_x,- k_y,- k_z) \label{eq:HamTdsymi}
\eneq
\beq
C_{2x} H(k_x, k_y, k_z) C_{2x}^{-1} = H(k_x, -k_y, -k_z)
\eneq
\beq
C_{2y} H(k_x, k_y, k_z) C_{2y}^{-1} = H(-k_x, k_y, -k_z)
\eneq
\beq
C_{2z} H(k_x, k_y, k_z) C_{2z}^{-1} = H(-k_x, -k_y, k_z)
\eneq
\beq
C_{4z}I H(k_x, k_y, k_z)  (C_{4z} I)^{-1} = H(k_y, -k_x, -k_z) 
\eneq
\beq
M_{1 10} H(k_x, k_y, k_z) M_{1 10}^{-1} = H(-k_y, -k_x, k_z)
\eneq
\beq
C_{3,111} H(k_x, k_y, k_z) C_{3, 111}^{-1}  =H(k_z, k_x, k_y) \label{eq:HamTdsymf}
\eneq

\subsection{Hamiltonian}
The most general  Hamiltonian consistent with Eqs.~(\ref{eq:HamTdsymi})-(\ref{eq:HamTdsymf}) is given -- to order $k^2$ -- by, 
%$H(k_x, k_y, k_z)  = H_0+  H_i k_i + H_{ij} k_i k_j $ where double index means summation. After imposing all the symmetries we find the following Hamiltonian:

\begin{eqnarray}
&H(k_x, k_y, k_z)=\left( A_0+ A k^2 \right) I_{4 \times 4}  + \nonumber \\
&+ C(k_x^2- k_y^2) \Gamma_1 + \frac{C}{\sqrt{3}} (2 k_z^2 - k_x^2- k_y^2) \Gamma_2 + \nonumber \\
& + E(k_x k_y \Gamma_3 + k_x k_z \Gamma_4 + k_y k_z \Gamma_5) + \nonumber  \\ 
& + D (k_x U_1 + k_y U_2 + 2 k_z U_3)  \label{kdotpzerofield-supp}
\end{eqnarray}
From here on, we ignore the overall energy shift by setting $A_0=0$.
Inversion symmetry is preserved when $D=0$. The matrices $\Gamma_{1\ldots5}, U_{1,2,3}$ are given by:
\begin{eqnarray}
&\Gamma_1 = \frac{1}{\sqrt3} (J_x^2- J_y^2) = \left(
\begin{array}{cccc}
 0 & 0 & 1 & 0 \\
 0 & 0 & 0 & 1 \\
 1 & 0 & 0 & 0 \\
 0 & 1 & 0 & 0 \\
\end{array}
\right) \nonumber \\ &
\Gamma_2 = \frac{1}{3} (2 J_z^2- J_x^2- J_y^2) = \left(
\begin{array}{cccc}
 1 & 0 & 0 & 0 \\
 0 & -1 & 0 & 0 \\
 0 & 0 & -1 & 0 \\
 0 & 0 & 0 & 1 \\
\end{array}
\right)\nonumber \\ &
\Gamma_3 =\frac{1}{\sqrt3}  \{J_x, J_y\} =\left(
\begin{array}{cccc}
 0 & 0 & -i & 0 \\
 0 & 0 & 0 & -i \\
 i & 0 & 0 & 0 \\
 0 & i & 0 & 0 \\
\end{array}
\right) \nonumber \\ &
\Gamma_4 =\frac{1}{\sqrt3}  \{J_x, J_z\} =\left(
\begin{array}{cccc}
 0 & 1 & 0 & 0 \\
 1 & 0 & 0 & 0 \\
 0 & 0 & 0 & -1 \\
 0 & 0 & -1 & 0 \\
\end{array}
\right) \nonumber \\ &
\Gamma_5=\frac{1}{\sqrt3} \{ J_y, J_z\}  = \left(
\begin{array}{cccc}
 0 & -i & 0 & 0 \\
 i & 0 & 0 & 0 \\
 0 & 0 & 0 & i \\
 0 & 0 & -i & 0 \\
\end{array}
\right) \nonumber \\ &
U_1=  \left(
\begin{array}{cccc}
 0 & 1 & 0 & \sqrt{3} \\
 1 & 0 & -\sqrt{3} & 0 \\
 0 & -\sqrt{3} & 0 & 1 \\
 \sqrt{3} & 0 & 1 & 0 \\
\end{array}
\right)   \nonumber \\ &
U_2 = \left(
\begin{array}{cccc}
 0 & i & 0 & -i \sqrt{3} \\
 -i & 0 & -i \sqrt{3} & 0 \\
 0 & i \sqrt{3} & 0 & i \\
 i \sqrt{3} & 0 & -i & 0 \\
\end{array}
\right) \nonumber \\ &
U_3 = \left(
\begin{array}{cccc}
 0 & 0 & -1 & 0 \\
 0 & 0 & 0 & 1 \\
 -1 & 0 & 0 & 0 \\
 0 & 1 & 0 & 0 \\
\end{array}
\right)
\end{eqnarray}
Notice the $\Gamma_{1\ldots5}$ are Clifford algebra matrices:
\beq
\{\Gamma_a, \Gamma_b\} = 2 \delta_{ab}
\eneq and $U_{1,2,3}$ can be written down in terms of the Clifford algebra matrices and their commutations (the full $su(4)$ algebra):
\beq
\Gamma_{ab} = \frac{1}{2i } [\Gamma_a, \Gamma_b]
\eneq
as:
\begin{eqnarray}
& U_1 = \sqrt{3} \Gamma_{15} - \Gamma_{25};\;\;\;\;\;\; U_2= - \sqrt{3} \Gamma_{14} - \Gamma_{24};\;\nonumber \\ &U_3 = \Gamma_{23}
\end{eqnarray}
The ab initio fit yields
\beq 
%%%%% notice that we formerly called A here B; I changed it to A to avoid later confusion with the magnetic field %%%%%
%%%%%% these are old units and old def of A
%\begin{equation}
%A=-28.3 a_0^2;\, C=40 a_0^2;\, D= 0.08a_0;\, E=-20a_0^2;
%\label{eq:abinitioparams}
%\end{equation}
%for HgTe:
%\begin{equation}
%A=-12.9 a_0^2;\, C=19.7 a_0^2;\, D= -.076a_0;\, E=-43a_0^2;
%\end{equation}
%and for InSb:
%\begin{equation}
%A=2.32 a_0^2;\, C=50.7a_0^2;\, D= -.019 a_0;\, E=-79a_0^2;
%\label{eq:abinitioparams2}
%\end{equation}
%%%%% these are new units
%\begin{equation}
%A=-1.46 {\rm eV \AA^2};\, C=8.20{\rm eV \AA^2};\,D=.08{\rm eV \AA};\,E=-5.4{\rm eV \AA^2};
%\label{eq:abinitioparams}
%\end{equation}
%for HgTe:
%\begin{equation}
%A=2.85 {\rm eV \AA^2};\, C=11.20{\rm eV \AA^2};\,D=.04{\rm eV \AA};\,E=-12{\rm eV \AA^2};
%\label{eq:abinitioparams1}
%\end{equation}
%and for InSb:
%\begin{equation}
%A=8.85 {\rm eV \AA^2};\, C=14.20{\rm eV \AA^2};\,D=.01{\rm eV \AA};\,E=-22{\rm eV \AA^2};
%\label{eq:abinitioparams2}
%\end{equation}
%\begin{equation}
\begin{array}{ccccc}
& A & C & D & E\\
{\rm GdPtBi} & -1.46 & 8.20 & 0.08 & -5.4\\
{\rm HgTe} & 2.85 & 11.20 & 0.04 & -12.0 \\
{\rm InSb} & 8.85 & 14.20 & 0.01 & -22.0 
\end{array}
\label{eq:abinitioparams}
\end{equation}
where $A$, $C$, and $E$ are in units of eV\AA$^2$ and $D$ is in units of eV\AA.
Along the $k_z$ axis, this structure is inverted, with the $J_z=\pm 3/2$ heavy bands dispersing upwards (electron-like) and the $J_z=\pm 1/2$ bands dispersing downwards (hole-like).

\subsection{Weyl Nodes}

When time-reversal symmetry is broken, a complicated phase diagram of Weyl points appears. The number of these Weyl points depends on the direction of the magnetic field, on the $g$-factors of the problem, and on whether the magnetic field couples differently to heavy and light holes. We can use the $k\cdot p$ model (\ref{kdotpzerofield-supp}), with the parameters (\ref{eq:abinitioparams}), to find the Weyl points in most of the interesting cases, and we expect it to be reliable. Since simulating $g$-factors is notoriously hard, we use an effective model for the Zeeman coupling where we add to the Hamiltonian in Eq.~(\ref{kdotpzerofield-supp}) the Zeeman term:
\beq
H_{Z}=\vec{B} \cdot \vec{J}
\label{eq:HZeeman}
\eneq where $J_x, J_y, J_z$ are the spin-$3/2$ matrices of Eqs.~(\ref{Jx}, \ref{Jy}, \ref{Jz}). We have assumed that the $g$-factor is positive, without loss of generality. We now analyze the spectrum as we align the magnetic field with high-symmetry crystallographic axes. 

\subsection{Field in $[001]$ direction}

At $\vec{k}=0$, the presence of the magnetic field splits the $3/2$ multiplet into four bands with energies $3/2B, 1/2B, - 1/2B, - 3/2B$; the first and fourth bands disperse upwards, while the second and third disperse downwards.

We first consider the spectrum when $\vec{k}$ is in the $[001]$ direction.
There are two Weyl nodes on the $001$ high symmetry axis between the $-3/2$ and $-1/2$ bands.
Since the band of energy $-3/2B$ at the $\Gamma$ point disperses upwards, while the bands of energy $1/2 B, -1/2 B$ disperse downwards, they will invariably intersect. The crossing between the $-1/2$ and $-3/2$ bands is protected, as the bands have different eigenvalue under $C_{2z}$ -- their eigenvalues are $\pm i$, as given by Eq~(\ref{c2z}).
There are two Weyls at $\vec{k} = (0,0, \pm \sqrt{ \frac{\sqrt{3} B}{4 C }})$. They are related by $C_{4z}I$, which means that they have opposite chirality due to the inversion operation (inversion changes the Weyl charge, TR leaves it invariant, rotations leave it invariant and mirrors -- the product of rotation times inversion -- change it.) They are shown in Fig.~\ref{fig:001easy}.

We move to analyze the physics between the $-3/2$ and $1/2$ bands (the $-3/2$ band disperses upwards, the $1/2$ band downwards and they start inverted).
Their crossings in the $[001]$ direction are avoided. However, we now give a topological proof why they must give rise to Weyl points elsewhere in momentum space.

Consider the plane $k_z=0$. It is a \emph{gapped} plane with respect to the bands $-3/2$, $1/2$. At the $\Gamma$ point, the $C_4I$ eigenvalues of these bands are $\exp(i 3\pi/4)$, $\exp(i 7\pi/4)$ respectively. The  $-3/2$ rests below the $1/2$ band at the $\Gamma$ point.

Away from the $\Gamma$ point, if we believe the $k\cdot p$ model correctly describes the physics, the bands at very large momentum (or $(\pi,\pi)$ on the lattice, which is infinity in the continuum)  are ordered in energy in the normal $1/2, -3/2$ way and so their $C_4$ eigenvalues are exactly the opposite of those at the $\Gamma$ point.

Now take the plane $k_z = \pi (=\infty) $. In this plane the bands have $C_{4z}I$ eigenvalues $\exp(i 7\pi/4), \exp(i 3\pi/4)$, ordered in energy. Hence, going from the $k_z=0$ to $k_z =\pi$ plane, the $C_{4z} I$ eigenvalue of the lower band has changed from $\exp(i 3\pi/4)$ to $\exp(i 7\pi/4)$. Ref~[\onlinecite{Fang12}] relates the eigenvalues of the $C_4$ operator of the occupied bands to the quantity $i^C$ where $C$ is the Chern number of the bands:
\begin{equation}
i^{C_{k_z=0} -C_{k_z = \infty} }=\frac{ \exp(i 3\pi/4)}{\exp(i 7\pi/4)} = -1
\label{eq:chern001}
\end{equation}
Hence, ${C_{k_z=0} -C_{k_z = \infty} } = 2$. Since both those planes are gapped with respect to the $-3/2$ and $1/2$ bands, we see that two (mod four) Weyl nodes have to reside between $k_z=0$ and $k_z = \infty$ planes. 

Suppose one of these points is at some $\vec{k} = (k_{x0}, k_{y0}, k_{z0})$. By $C_{2z}$ symmetry, its partner is at $(-k_{x0}, -k_{y0}, k_{z0})$.  However, unless $k_{x0}=0$ or $k_{y0}=0$, the product $C_{2x}T$, which remains a symmetry in the presence of a magnetic field (with a magnetic field in the $[001]$ direction, $C_{2x}$ flips the field, but time-reversal flips it back), implies two additional Weyl points between $k_z=0$ and $k_z = \infty$ planes at $(\mp k_{x0}, \pm k_{y0}, k_{z0})$. Thus, the two Weyl points required by the Chern number argument in Eq~(\ref{eq:chern001}) must lie in either the $k_x=0$ or $k_y=0$ plane. Figure \ref{fig:001hard} shows the pair of Weyls in the $k_x=0$ plane.

Suppose they lie in the $k_x=0$ plane, at $(0, k_{y0}, k_{z0})$.
Then, by $C_{4z}I$ symmetry, there must be two more Weyl points at $(\pm k_{y0},0, -k_{z0})$ (which could have been derived by applying the same Chern number argument in Eq~(\ref{eq:chern001}) to the $k_z=0$ and $k_z=-\infty$ planes).

In summary, we have proven that (at least) six Weyl points must exist  -- two on a  high symmetry line and two on high-symmetry planes.
In general, additional Weyl points might exist at generic points in the Brillioun zone.
The combined presence of $C_{4z}I$ and $C_{2x}T$ implies that these Weyl points come in sets of eight. Thus, the total number of Weyl points that can exist is $6+8k$, where $k\in\mathbb{Z}$. This is shown in Table I.

Due to the $C_3$ symmetry of the crystal, an identical argument goes through for $\vec{B}$ aligned with the $100$ and $010$ directions as well.

\subsection{Field in $[110]$ direction}
With $\vec{B}$ in the $[110]$ direction, it is convenient to work in the basis of eigenstates of 
\begin{equation}
J_x+J_y\equiv J_{110}. 
\end{equation}
At the $\Gamma$ point, eigenstates of the Hamiltonian coincide with eigenstates of $J_{110}$. As before, the bands with $J_{110}$ eigenvalue $\pm 3/2$ disperse upwards, and the bands with eigenvalue $\pm 1/2$ disperse downwards.

With the magnetic field in this direction, the only remaining symmetries of the Hamiltonian are the mirror symmetry $M_{110}$, and the combined symmetry $C_{2z}T$. The points $\vec{k}=(k,-k,k_z)$ are fixed by the action of $M_{110}$, while the points $\vec{k}=(k_x,k_y,0)$ are fixed by $C_{2z}T$; these two subspaces are the high-symmetry planes we will analyze. (The intersection line $\vec{k}=(k,-k,0)$ of these two planes is invariant under $M_{110}C_{2z}T$. However, as an anti-unitary symmetry that squares to $+1$, it does not permit band crossings on a line without fine-tuning, for the following reason: a generic two-band Hamiltonian on the line takes the form $H(k)=d(k)\cdot\sigma$. An anti-unitary symmetry that squares to $+1$ can be represented by $K$, the complex conjugation operator. The requirement $[H(k),K]=0$ implies $d_y(k)=0$. Thus, $H(k)$ is described by two functions, $d_{x,z}$, which are a function of one variable, $k$. Without fine-tuning, it is not generically true that there exists a $k_0$ where $d_x(k_0) = d_z(k_0) = 0$.)

We will start by analyzing the plane fixed by $M_{110}$. Along this plane the energy eigenstates are $M_{110}$ eigenstates. Going away from the $\Gamma$ point, the bands with $J_{110}$ eigenvalues $(3/2,1/2,-1/2,-3/2)$ have, respectively, the mirror eigenvalues $(i,-i,i,-i)$. Based on the direction of the band dispersion, we then expect the $J_{110}=-1/2$ band and the $J_{110}=-3/2$ bands to cross in the high symmetry plane due to their differing mirror eigenvalues.

It is not hard to show that such crossings produce \emph{line nodes}. In the vicinity of the crossing, we can represent the mirror symmetry, which squares to $-1$, as $M_{110}=i\sigma_y$. The effective Hamiltonian on the high symmetry plane takes the form
\begin{equation}
H_{eff}(k,-k,k_z)=d_i(k,-k,k_z)\sigma_i.
\end{equation}
Imposing the symmetry $M_{110}H_{eff}(k,-k,k_z)M_{110}^{-1}=H_{eff}(k,-k,k_z)$ requires
\begin{equation}
H_{eff}(k,-k,k_z)=d_y(k,-k,k_z)\sigma_y.
\end{equation}
Since there is one parameter $d_y$ and two momenta $(k,k_z)$, any point where $d_y(k,-k,k_z)=0$ is generically part of a line node where $d_y(k,k_z)=0$.

Using the ab initio parameters (\ref{eq:abinitioparams}), we are guaranteed such a point. Furthermore, when $D=0$, so that inversion symmetry is present, we have two such line nodes at inversion-symmetric points. We expect this to hold true for small values of $D$ as well, and for the physically relevant situation $D=0.08$, we do, in fact, observe two line nodes, which are shown in Fig.~\ref{fig:110}. However, as inversion symmetry breaking is increased to unphysical values, the bands tilt so that they only cross on one side of the origin and only one line node exists.

% The situation is a bit different for $M_{110}T$. This fixes the high symmetry line $\vec{k}=(k,-k,0)$, along which it can be represented by $K$ in a two-band effective model. As such, the effective Hamiltonian is given by
% \begin{equation}
% H_{eff}(k)=d_x(k)\sigma_x+d_z(k)\sigma_z.
% \end{equation}
% Since this has two parameters and one momenta, There is generically no band crossing, and no Weyl protected by this symmetry.

Let us now examine the high symmetry $(k_x,k_y,0)$ plane. This plane is fixed by $C_{2z}T$, which is antiunitary and squares to $+1$. Assuming there exists a crossing on this plane, we can write an effective two-band Hamiltonian, and represent $C_{2z}T$ by $K$. An analysis similar to the previous section shows that after enforcing this symmetry, the Hamiltonian consists of two parameters, which are a function of two momenta. Hence, on this plane, there can generically be isolated band crossings in the $(k_x,k_y,0)$ plane. 
The $M_{110}$ symmetry requires that a putative Weyl point at $(k_x,k_y,0)$ have a partner Weyl at $(-k_y,-k_x,0)$.

If the line nodes in the $(k,-k,k_z)$ plane persist to $k_z=0$, they will intersect the $(k_x,k_y,0)$ plane, twice each. By tuning the parameters in the Hamiltonian (for instance the degree of inversion symmetry breaking), we can change the number of line nodes. 
Hence, the line nodes can create pairs of band degeneracies in the plane $(k_x,k_y,0)$, with the number of pairs determined by the ab initio parameters.
Using the particular ab initio values (\ref{eq:abinitioparams}), we observe four band crossings in the $(k_x,k_y,0)$ plane, and all four lie on the line nodes.

Similar arguments go through when considering the $3/2$ and $1/2$ band, which also have differing mirror eigenvalues. We find from the $k\cdot p$ Hamiltonian that there is also a pair of line nodes between these bands, however, using the ab initio parameters, this pair sits much farther from the $\Gamma$ point in the Brillouin zone; whether or not this pair is an artifact of the $k\cdot p$ expansion would have to be confirmed by further ab initio calculations.
%\textbf{Such line nodes could be artifacts of the $k\cdot p$ expansion; further ab-initio calculations should be done to confirm the existence of these lines.}
%There cannot, however, be an odd number of crossings in this plane: we know that Weyls must come in pairs, and so a single Weyl node on this plane must be related to another Weyl node not in the plane by symmetry. However, all remaining symmetries of the Hamiltonian leave the plane $(k_x,k_y,0)$ invariant; this is a contradiction. We thus conclude that there are an even number of nodes on this plane. Finally, we find that in our model, the only such nodes correspond to the intersection of the mirror line nodes with the plane.

Finally, we find four additional Weyl points near the $(k,k,k_z)$ plane. We prove their existence in the inversion-symmetric case $D=0$, where they lie exactly on this plane; the nontrivial Chern number associated with the Weyl points guarantees that they survive small inversion breaking. 
When $D=0$, there is an additional unitary symmetry, $C_{2,110}$, and antiunitary symmetry, $C_{2,1\bar{1}0}T$, which satisfy,
\begin{equation} 
C_{2,110}^2=-1; (C_{1\bar{1}0}T)^2=1; \{ C_{2,110},C_{2,1\bar{1}0}T\}=0
\label{eq:B110symmrels}
\end{equation}
Both of these symmetries leave invariant the line $\vec{k}=(k,k,0)$; the former also leaves invariant the plane $\vec{k}=(k,k,k_z)$. 

We start by considering the special value of parameters, $E=2C$ in Eq~(\ref{kdotpzerofield-supp}). On the line $k_x=k_y, k_z=0$, the Hamiltonian reduces to:
\begin{equation}
H(\frac{k_{110}}{2},\frac{k_{110}}{2},0) = A\frac{k_{110}^2}{2}\mathbb{I} +\frac{Ck_{110}^2}{2}\left( -\frac{\Gamma_2}{\sqrt{3}}+\Gamma_3\right)
\label{eq:HE2C}
\end{equation}
where we define $k_{110(1\bar{1}0)} = \left(k_x +(-)k_y\right)/2$.
The Hamiltonian (\ref{eq:HE2C}) commutes with $J_x+J_y$.
Thus, when the Zeeman term (\ref{eq:HZeeman}) is added, the four bands can be labelled not only by their $C_{2,110}$ eigenvalue, but by their eigenvalue under $(J_x+J_y)/\sqrt{2}$. The bands with eigenvalues $\pm 1/2$ both cross the band with eigenvalue $-3/2$. We now analyze each of these crossings to show which are protected as we move away from the $E=2C$ point -- if the band crossings are Weyl points, they will persist for small perturbations, but we want to know whether they persist for a large change in parameter space, where $E$ and $C$ are given by the parameters (\ref{eq:abinitioparams}).

We first consider the crossings between the $-1/2$ and $-3/2$ bands. An effective Hamiltonian for these two bands is given by $H_{\rm eff}^{-\frac{3}{2},-\frac{1}{2}} = d_i(k_{110},k_{1\bar{1}0},k_z)\sigma_i$, for some functions $d_i$. In this effective space, we can choose $C_{2,110}$ to be represented by $i\sigma_y$ and $C_{1\bar{1}0}T$ by $\sigma_zK$, where $K$ is the complex conjugation operator; this choice is consistent Eq~(\ref{eq:B110symmrels}) and consistent with the fact that these bands have distinct eigenvalues $\pm i$ under $C_{2,110}$. Enforcing $C_{2,110}$ yields
\begin{align}
d_{x,z}(k_{110},k_{1\bar{1}0},k_z) &= -d_{x,z}(k_{110},-k_{1\bar{1}0},-k_z)\nonumber\\
d_{y}(k_{110},k_{1\bar{1}0},k_z) &= d_{y}(k_{110},-k_{1\bar{1}0},-k_z),
\end{align}
while enforcing $C_{1\bar{1}0}T$ yields,
\begin{align}
d_{x}(k_{110},k_{1\bar{1}0},k_z) &= -d_{x}(k_{110},-k_{1\bar{1}0},k_z)\nonumber\\
d_{y}(k_{110},k_{1\bar{1}0},k_z) &= d_{y}(k_{110},-k_{1\bar{1}0},k_z)\nonumber\\
d_{z}(k_{110},k_{1\bar{1}0},k_z) &= d_{z}(k_{110},-k_{1\bar{1}0},k_z)
\end{align}
To linear order in $k$, these sets of constraints yield $d_x\propto k_{1,\bar{1}0},d_z \propto k_z,d_y\propto k_{110}+d_0$, for some constant $d_0$ that determines the location of the band crossing. Hence, the band crossings between these bands are two single Weyl points. 
When $E$ is tuned away from $2C$, the Weyl points will remain on the line $(k,k,0)$ (Weyl points off this line come in sets of four or eight). 
%Thus, they can only annihilate at the $\Gamma$ point or at $k\rightarrow \infty$ (which corresponds to $k=\pi$ in a tight-binding model). It is clear that the $\Gamma$ point is gapped for all values of $C$ and $E$; hence, we rule out the possibility of the Weyl points annihilating there. We now consider the possibility that they annihilate at $\infty$.
Thus, to determine in what parameter regime the Weyl points persist, we need only diagonalize $H(k,k,0)+(J_x+J_y)B/\sqrt{2}$ and check for crossings between the $-3/2$ and $-1/2$ bands. This Hamiltonian has eigenvalues,
\begin{equation}
\eta\frac{B}{2} \pm \sqrt{B^2+\eta\frac{B}{\sqrt{3}}k^2\left(2C+3E\right)+\frac{4}{3}C^2k^4+E^2k^4}
\end{equation}
where $\eta=\pm 1$ and we have omitted the shift $2A k^2$ in all band energies.
The $-3/2$ and $-1/2$ bands intersect only when $E(2C+E)>0$. Thus, for $C>0$, the Weyl points exist only when $E>0$ or $E<-2C$ (when $C<0$, they exist when $E<0$ or $E>2|C|$); at the transition points, the Weyls annihilate at $k\rightarrow \infty$. Referring to the ab initio values in (\ref{eq:abinitioparams}), we see that in the regime of interest for this particular material, the Weyl points between these two bands do not exist. We have confirmed this analysis of the $k\cdot p$ model numerically.

We next consider the crossing between the $1/2$ and $-3/2$ bands.
Instead of applying a symmetry argument similar to that in the previous paragraph, we directly expand $H(k_x,k_y,k_z)+B(J_x+J_y)/\sqrt{2}$ about the desired band crossing at $k_{110}=3^{1/4}\sqrt{B/C}, k_{1\bar{1}0}=k_z=0$ and project onto the $1/2$ and $-3/2$ bands to compute the $2\times 2$ effective Hamiltonian, $ H_{\rm eff}^{-\frac{3}{2},\frac{1}{2}} = c_0\mathbb{I} + c_i\sigma_i$,
where
\begin{align}
c_0 &= \frac{(A\sqrt{3}-3C)B}{2C}+\frac{\sqrt{BC}}{3^{1/4}}\delta_{110}+\frac{C}{\sqrt{3}}\left(k_{1\bar{1}0}^2+2k_z^2\right)  \nonumber\\
c_x &= -\frac{3^{1/4}\sqrt{BC}}{\sqrt{2}}\delta_{110}+\frac{C}{\sqrt{2}}k_{1\bar{1}0}^2 -2Ck_{1\bar{1}0}k_z \nonumber\\
c_y &= \frac{3^{1/4}\sqrt{BC}}{\sqrt{2}}\delta_{110}-\frac{C}{\sqrt{2}}k_{1\bar{1}0}^2 -2Ck_{1\bar{1}0}k_z \nonumber\\
c_z &= -\frac{\sqrt{BC}}{3^{1/4}}\delta_{110} - \frac{C}{\sqrt{3}}\left(k_{1\bar{1}0}^2-4k_{z}^2\right)
\end{align}
where $\delta_{110}\equiv k_{110}-3^{1/4}\sqrt{B/C}$. Defining a rotated set of Pauli matrices
$\tilde{\sigma}_i=\mathcal{O}_{ij}\sigma_j$, where 
\begin{equation}\mathcal{O}=\begin{pmatrix}1/\sqrt{2}&1/\sqrt{2}&0\\1/(2\sqrt{2})&-1/(2\sqrt{2})&-\sqrt{3}/2 \\ \sqrt{3}/(2\sqrt{2}) & -\sqrt{3}/(2\sqrt{2}) & 1/2 \end{pmatrix},\end{equation}
we find
$ H_{\rm eff}^{-\frac{3}{2},\frac{1}{2}} = \tilde{c}_i\tilde{\sigma}_i $, where 
\begin{align}
\tilde{c}_x &= -2\sqrt{2}Ck_{1\bar{1}0}k_z \nonumber\\
\tilde{c}_y &= C(k_{1\bar{1}0}^2-2k_z^2) \nonumber\\
\tilde{c}_z &= \tilde{\delta}_{110}
\end{align}
where $\tilde{\delta}_{110} \equiv -2(3^{-1/4})\sqrt{BC}\delta_{110}+C(k_{1\bar{1}0}^2+2k_z^2)/\sqrt{3}$ is a shift and rescaling of the $\delta_{110}$ axis. In this form, it is evident that $H_{\rm eff}^{-\frac{3}{2},\frac{1}{2}}$ describes a double Weyl point. When $E$ is tuned away from its special value of $2C$, this double Weyl, and its partner at $k_{110}=-3^{1/4}\sqrt{B/C}, k_{1\bar{1}0}=k_z=0$, split into four single Weyl points, which are restricted to lie in the $(k_{110},k_z)$ plane (because Weyl points off this plane come in sets of eight). The single Weyl points can pairwise annihilate only on the $k_z$ axis. Thus, if the $k_z$ axis is gapped for all value of $E$ at fixed $C$, then these four single Weyl points persist through the large parameter change from $E=2C$ to $E$ and $C$ described by (\ref{eq:abinitioparams}). The Hamiltonian along the $k_z$ axis, including the Zeeman term, is given by 
$H(0,0,k_z)+(J_x+J_y)B/\sqrt{2}$, which has eigenvalues $B\left(-1/2+x^2/\sqrt{3}\pm \sqrt{1+\frac{2}{\sqrt{3}}x^2+\frac{4}{3}x^4}\right)$ and $B\left(1/2+x^2/\sqrt{3}\pm \sqrt{1-\frac{2}{\sqrt{3}}x^2+\frac{4}{3}x^4}\right)$, where $x=\sqrt{C/B}k_z$. It is straightforward to check that these eigenvalues are distinct for all real values of $x$. Thus, we have shown that the two double Weyl points that describe the crossings between the $-3/2$ and $+1/2$ bands at the special parameter value $E=2C$ split into four single Weyl points, which survive as $E$ and $C$ are varied to the values in (\ref{eq:abinitioparams}). 

%%%%%% symmetry argument for the 1/2, -3/2 bands %%%%%%%%
%Since these bands have the same eigenvalue under $C_{2,110}$, we can choose $C_{2,110}$ to be represented by $i\mathbb{I}$ and $C_{2,1\bar{1}0}T$ by $K$; this is also consistent with Eq~(\ref{eq:B110symmrels}). Again, we write an effective Hamiltonian, $H_{\rm eff} = c_i(k_{110},k_{1\bar{1}0,k_z})$. Enforcing $C_{2,110}$ yields the constraints,
%\begin{align} c_{x,y,z}(k_{110},k_{1\bar{1}0},k_z) &= c_{x,y,z}(k_{110},-k_{1\bar{1}0},-k_z),\end{align}while enforcing $C_{2,1\bar{1}0}T$ yields,
%\begin{align}c_{x,z}(k_{110},k_{1\bar{1}0},k_z) &= c_{x,z}(k_{110},-k_{1\bar{1}0},k_z)\nonumber\\  c_{y}(k_{110},k_{1\bar{1}0},k_z) &= -c_{y}(k_{110},-k_{1\bar{1}0},k_z)\end{align}
%Thus, to leading order in $k$, 
%\begin{align}c_{x,z}&=\delta+\alpha k_{110}+ \beta_{x,z} k_{1\bar{1}0}^2+\gamma_{x,z} k_z^2\nonumber \\ c_y&=\epsilon k_zk_{1\bar{1}0} \end{align}
%for some coefficients $\alpha,\beta_{x,z},\gamma_{x,z},\delta,\epsilon$; we have used the fact that the band crossing occurs on the $k_{110}$ axis.

Restoring $D=0.08\neq0$, we argue that, while these nodes may move off the symmetry plane, they do not annihilate. This is because their opposite chirality partners are generically far away in the Brillouin zone. 
%\textbf{This is confirmed via the ab-initio calculation.}

Lastly, due to the symmetries of the crystal, a completely analogous analysis holds for $\vec{B}$ aligned with any of the $[101],[011],[\bar{1}10],[\bar{1}01]$, or $[0\bar{1}0]$ directions.

% \subsection{Field in $1\bar{1}0$ direction}

% Observe that $C_{2x} (J_x - J_y) C_{2x}^{-1} = J_x + J_y$. It follows that the spectrum when $\vec{B} || [1\bar{1} 0]$ is identical to that when $\vec{B} || [110]$ and the same analysis of Weyl points and nodes follows.

% We could also have seen this by noting that when $\vec{B} || [1\bar{1} 0]$, the total Hamiltonian has $M_{110} \equiv C_{2x}M_{1\bar{1}0} C_{2x}^{-1}$ symmetry, satisfying 
% \begin{equation} M_{110} H(k_x,k_y,k_z) M_{110}^{-1} = H(k_y,k_x,k_z)\end{equation}
% (the $\vec{J}\cdot\vec{B}$ term is invariant.) Hence, the analysis of the previous section along the $(k, -k, k_z)$ plane applies here to the $(k,k,k_z)$ plane. 

\subsection{Field in $[111]$ direction}

Define $J_{111} = \frac{1}{\sqrt{3}}(J_x + J_y + J_z)$. Then 
\begin{align}
C_{3,111}J_{111}C_{3,111}^{-1}&=J_{111}.
\end{align}
Thus, when $\vec{B}||[111]$, eigenstates of $J_{111}$ are simultaneously eigenstates of $C_{3,111}$. In particular, the states with $J_{111}$ eigenvalues $3/2,-3/2,1/2,-1/2$ have $C_{3,111}$ eigenvalues $-1, -1,(1-i\sqrt{3})/2, (1+i\sqrt{3})/2$, respectively.

At $\vec{k}=0$, the bands corresponding to $\pm 3/2$ curve downwards, while the bands corresponding to $\pm 1/2$ curve upwards (notice that along the $\hat{x}, \hat{y}, \hat{z}$ directions, along the $\vec{k}|| \vec{B}$ axis, the opposite is true, that is, the $\pm 3/2$ bands curve upwards.) Hence, there are possible crossings between the 3/2 band and the $\pm 1/2$ bands (while the $-3/2$ band has no potential crossings.) Along the high-symmetry axis $k_x = k_y=k_z$, these bands all have distinct eigenvalues of $C_{3,111}$; hence, these crossings are protected, and we see four Weyl points.

At the inversion symmetric point ($D=0$), these Weyl points are exactly at $\vec{k} = \pm \sqrt{B/E}(1,1,1)$ and $\pm \sqrt{B/(2E)}(1,1,1)$, where $\vec{B} = B(1,1,1)$. When $D\neq 0$, we can also find an analytical expression for the Weyl points along this axis, but it is not elucidating. These four Weyls are shown in Fig.~\ref{fig:111}.
%Even without inversion symmetry, the points occur in pairs $\pm (k,k,k)$ because along this axis the Hamiltonian has the symmetry 
%\begin{equation} C_{4z}IKH(k,k,k) K^{-1}I^{-1}C_{4z}^{-1}= H(-k,-k,-k), \end{equation} while $J_x+J_y+J_z$ is invariant under this operation.

Notice that if there is an additional Weyl node at some point $(k_{x0},k_{y0},k_{z0})$, then the combination of $C_{3,111}$ and $C_{4z}I$ dictates that there are five more symmetry-related Weyls. Hence, the total number of Weyl nodes is $4+6k$; numerical diagonalization of the $k\cdot p$ Hamiltonian in this case indicates that $k=1$ for the ab initio parameters~(\ref{eq:abinitioparams}).

%%%%%%
%%%%%%
%%%%%%
%%%%%%
%%%%%%
%%%%%%

\subsection{Proof that Weyl points exist for all directions of $\mathbf{B}$}
We now prove that for any generic material parameters, Weyl points exist in all directions of the magnetic field when inversion symmetry is present; the topological nature of the Weyl points ensures that they will persist for small inversion symmetry-breaking.
For arbitrary $\mathbf{B}$, consider the Hamiltonian $H(\mathbf{k})+H_Z \equiv H(\mathbf{k}) + \mathbf{J}\cdot\mathbf{B}$, where $H(\mathbf{k})$ is defined in Eq~(\ref{kdotpzerofield-supp}); then set $D=0$ to preserve inversion symmetry and, without loss of generality, set $A=A_0=0$, which fixes the zero of energy.
The Hamiltonian is now a function of the parameters $E$ and $C$, as well as of $\mathbf{k}$ and $\mathbf{B}$.
Now consider the fine-tuned point in parameter space, $E=2C$: with this choice of parameters, $[ H(\mathbf{k}), \mathbf{k}\cdot \mathbf{J}]=0$, and, in particular, $[ H(k\mathbf{B}), \mathbf{B}\cdot \mathbf{J}]=0$, rendering the Hamiltonian exactly solvable along the axis $\mathbf{k} \parallel \mathbf{B}$.
The eigenvalues of $H(k\mathbf{B}) + \mathbf{J}\cdot\mathbf{B}$ are given by $ \frac{1}{6} \left( -4\sqrt{3}Ck^2B^2 \pm 3B \right)$ and $\frac{1}{6} \left( 4\sqrt{3}Ck^2B^2 \pm 9B \right)$.
For finite $B$, Weyl points exist at $k=\pm 3^{1/4}/\sqrt{4C}$ and $k=\pm 3^{1/4}/\sqrt{2C}$, as shown below. The band crossings are protected because all four bands have different eigenvalues of $\mathbf{B}\cdot \mathbf{J}$.
\begin{figure}[h]
\centering
\includegraphics[width=0.4\textwidth]{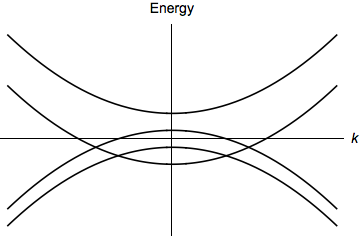}
% \label{fig:WeylsFineTuned} no Fig labels in this formatj
\caption{Schematic depiction of the four Weyl points that exist along the direction $\mathbf{k}\parallel \mathbf{B}$ when $E=2C$.}
\end{figure}
As $E$ is varied away from $2C$, the four Weyl points move out of the $\mathbf{k}\parallel \mathbf{B}$ plane, to generic positions $\pm \mathbf{k}_{1,2}$, consistent with inversion symmetry.
Notice that the Weyl points at $\pm \mathbf{k}_1$ cannot annihilate each other (nor can the points at $\pm \mathbf{k}_2$) because to do so while enforcing inversion symmetry would require two bands to become degenerate at the origin; however, for finite $\mathbf{B}$, the Hamiltonian has four distinct eigenvalues at the origin (it is equal to $\mathbf{J}\cdot\mathbf{B}$, which has eigenvalues $\pm \frac{1}{2}B$, $\pm \frac{3}{2}B$).
Thus, the Weyl points can only annihilate if there exist values of $E$ and $C$ such that $\mathbf{k}_1=\pm \mathbf{k}_2$.
%Thus, for the Weyl points to annihilate, there must exist $E_*$ and $C_*$, such that $\mathbf{k}_1|_{E_*,C_*} = \pm \mathbf{k}_2|_{E_*,C_*} \neq \mathbf{0}$.

If $\mathbf{k}_1$ is near $\pm \mathbf{k}_2$, the effective Hamiltonian of the two nearby Weyl points requires only three bands, i.e., it is a $3\times 3$ Hermitian matrix, which is described by eight functions, $d_i(\mathbf{k}), i=1,\dots, 8$, which are determined by $E$ and $C$. The band crossing occurs when $d_i(\mathbf{k}) = 0$ for all $i$.
Generically, a system of eight equations involving five variables ($E$, $C$, and $k_{x,y,z}$) has no solutions.
If solutions exist, then they exist as isolated points, $(E_*, C_*, \mathbf{k}_*)$.

We now argue that these Weyl points persist for physical values of the material parameters. Given a particular material, described by generic values $E_1 \neq 2C_1$, consider tuning $E$ and $C$ from $(E,C)=(2C_0,C_0)$, for arbitrary $C_0\neq 0$, to $(E_1,C_1)$. We can always choose a path in parameter space  that avoids the isolated points $(E_*,C_*)$ which permit three-band crossings. This leaves no opportunity for the Weyl points that exist at $(2C_0,C_0)$ to annihilate.
Hence, for arbitrary $B$ and generic values of $E$ and $C$, at least four Weyl points exist.

%%%%%%
%%%%%%
%%%%%%
%%%%%%
%%%%%%
%%%%%%

\subsection{Numerical Observations}
We confirmed the above analysis for GdPtBi, HgTe and InSb by numerically diagonalizing the $k\cdot p$ Hamiltonian, using the parameters in Eq~(\ref{eq:abinitioparams}). Surprisingly, this procedure showed that there exist Weyl nodes for the magnetic field in \emph{any} direction; the field-induced Weyls are surprisingly universal. Furthermore, we find that while all of the Weyl nodes on the high symmetry planes mentioned in the previous sections were of Type I, some of the Weyl fermions that emerge in the numerics are the Type II Weyls introduced in Ref~\onlinecite{Soluyanov15}. We now list some mention some specific results for GdPtBi. Focusing on the middle two bands and for magnetic fields of unit magnitude in the $[11x]$ direction, we find that there are at least four Weyl nodes for any $x\in[0,1]$, and that for $x=0.4121$, two of these nodes transition from Type I to Type II. These critical Weyls are located at $\mathbf{k}=(-0.0461,-0.302,0.244)$ and $\mathbf{k}=(0.302,0.0461,-0.244)$ (The symmetry in location is due to $M_{110}C_{2z}T$%=C_{1\bar{1}0}K$
, which preserves any magnetic field of the form $(B,B,B_z)$).
The energy of these crossings is $E=-0.225$. The two Type I Weyls are located at $\mathbf{k}=(0.318,0.312,0.330)$ and $\mathbf{k}=(-0.312,-0.318,-0.330)$, with energies $E=-0.03$.

\begin{figure*}\centering
\subfloat[\label{fig:001easy}]{
\includegraphics[width=0.4\textwidth]{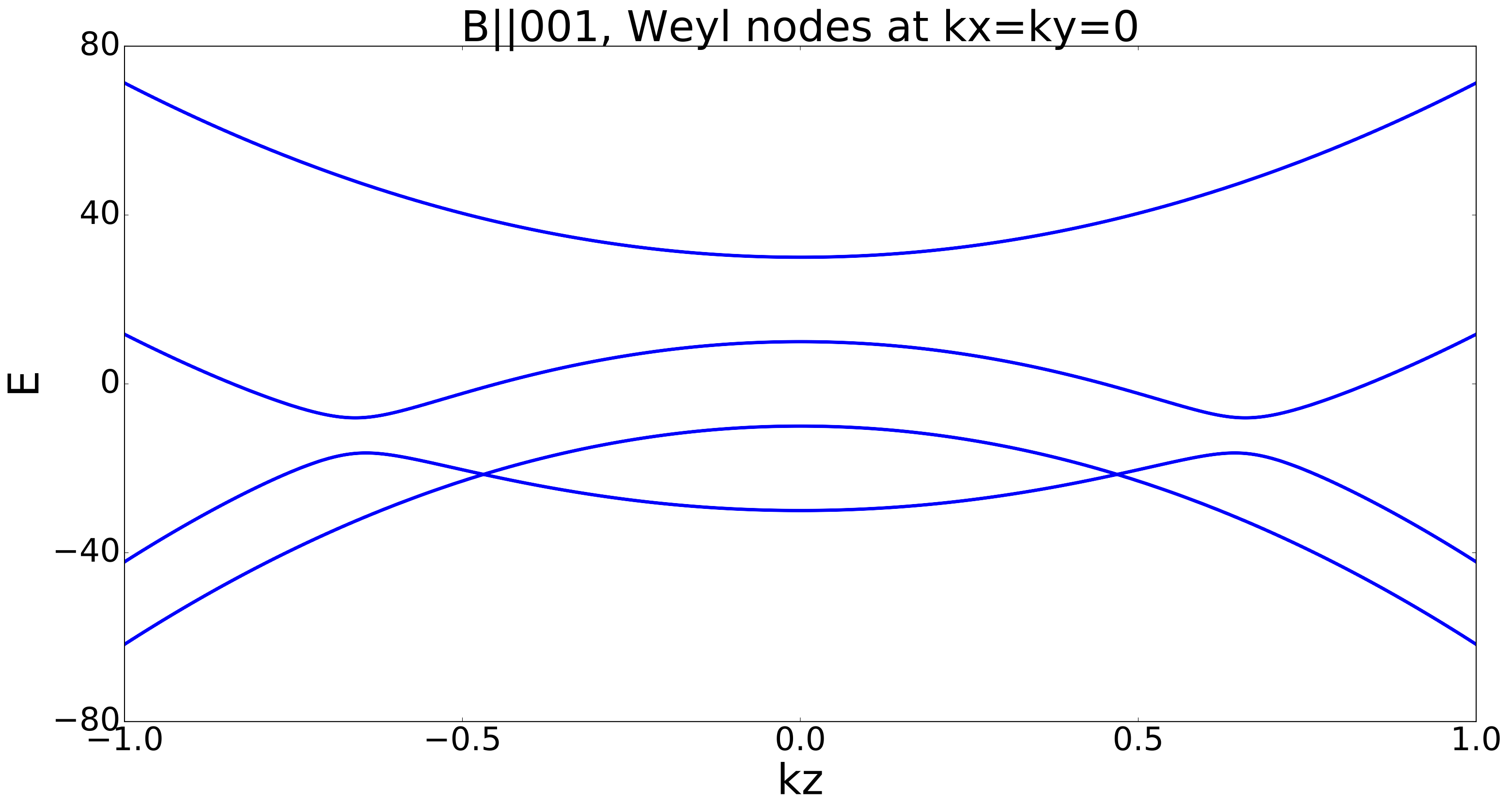}}
\subfloat[\label{fig:001hard}]{
\includegraphics[width=0.4\textwidth]{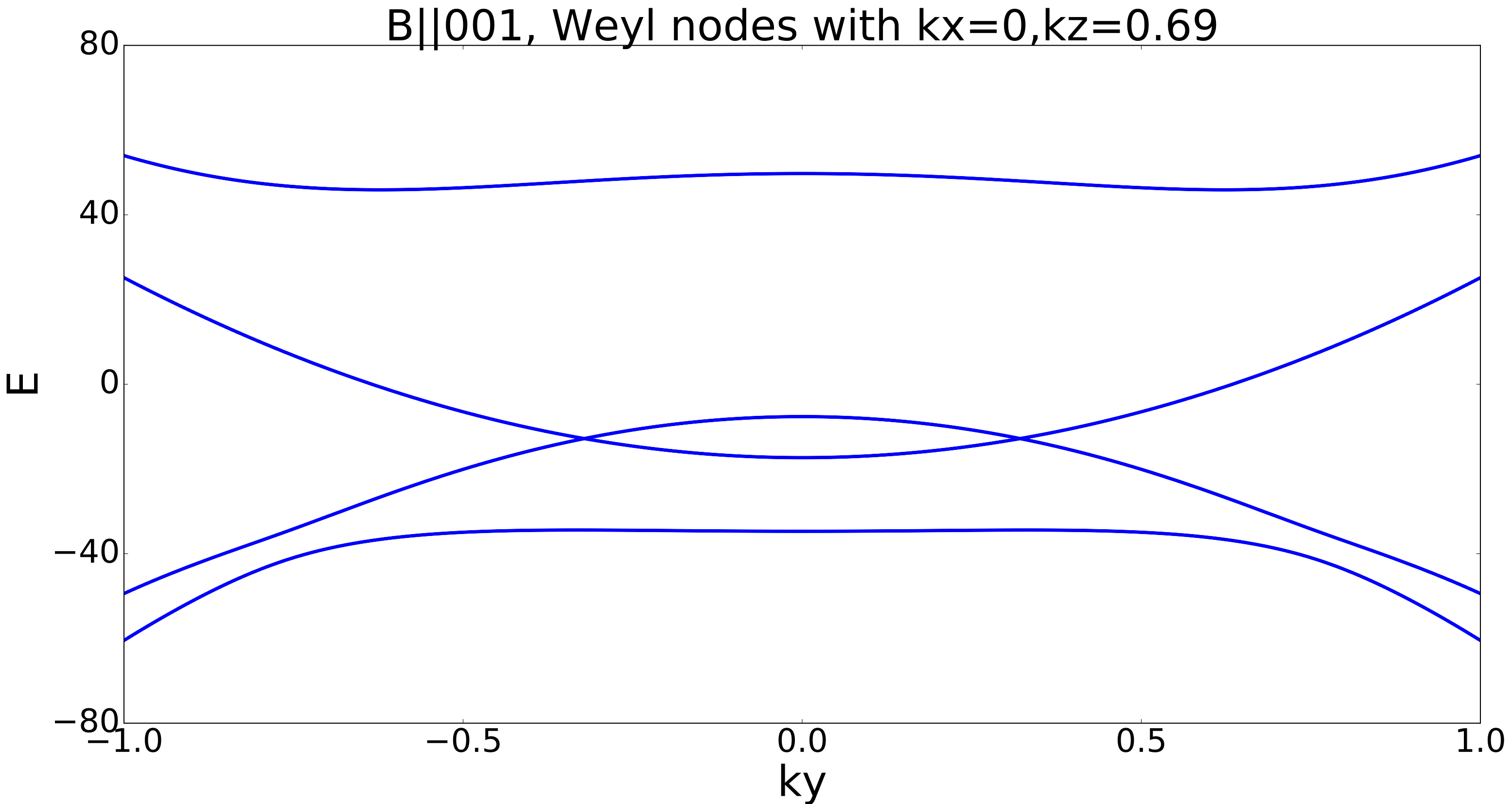}}
\caption{(a) shows two Weyl nodes along the line $(0,0,k_z)$ with $\vec{B}=(0,0,20)$. In the figure we have increased $D$ by a factor of $40$ in order to resolve the avoided crossings. The Weyl nodes are type I, and are described by an effective two-band Hamiltonian with $\vec{v}=\pm(0,0,4.85)$ and $A=\left(
\begin{array}{ccc}
-9.39 & 0 & 0 \\
0 & -9.23 & 0 \\
0 & 0 & \pm 42.98
\end{array}
\right)
$.
(b) shows the non-trivial Weyl nodes in the $k_x=0$ plane, for the same values of the field and material parameters ($D=5$) as in the previous figure. The linearized Hamiltonian for the rightmost Weyl has (for $D=0.08$) $\vec{v}=(0,-0.24,6.12)$ and $
A=\left(
\begin{array}{ccc}
0 & 1 & 0 \\
 3.23 & 0 & 2.94 \\
 19.55& 0& 57.57 \\
\end{array}
\right)
$. This is a Type I Weyl. The other is related to it by $C_{2z}$.}
\end{figure*}
% \begin{figure*}
% \centering
% \includegraphics[width=0.4\textwidth]{B111.png}
% \caption{}
% \end{figure*}

\begin{figure*}
\centering
\subfloat[\label{fig:110}]{
\includegraphics[width=0.4\textwidth]{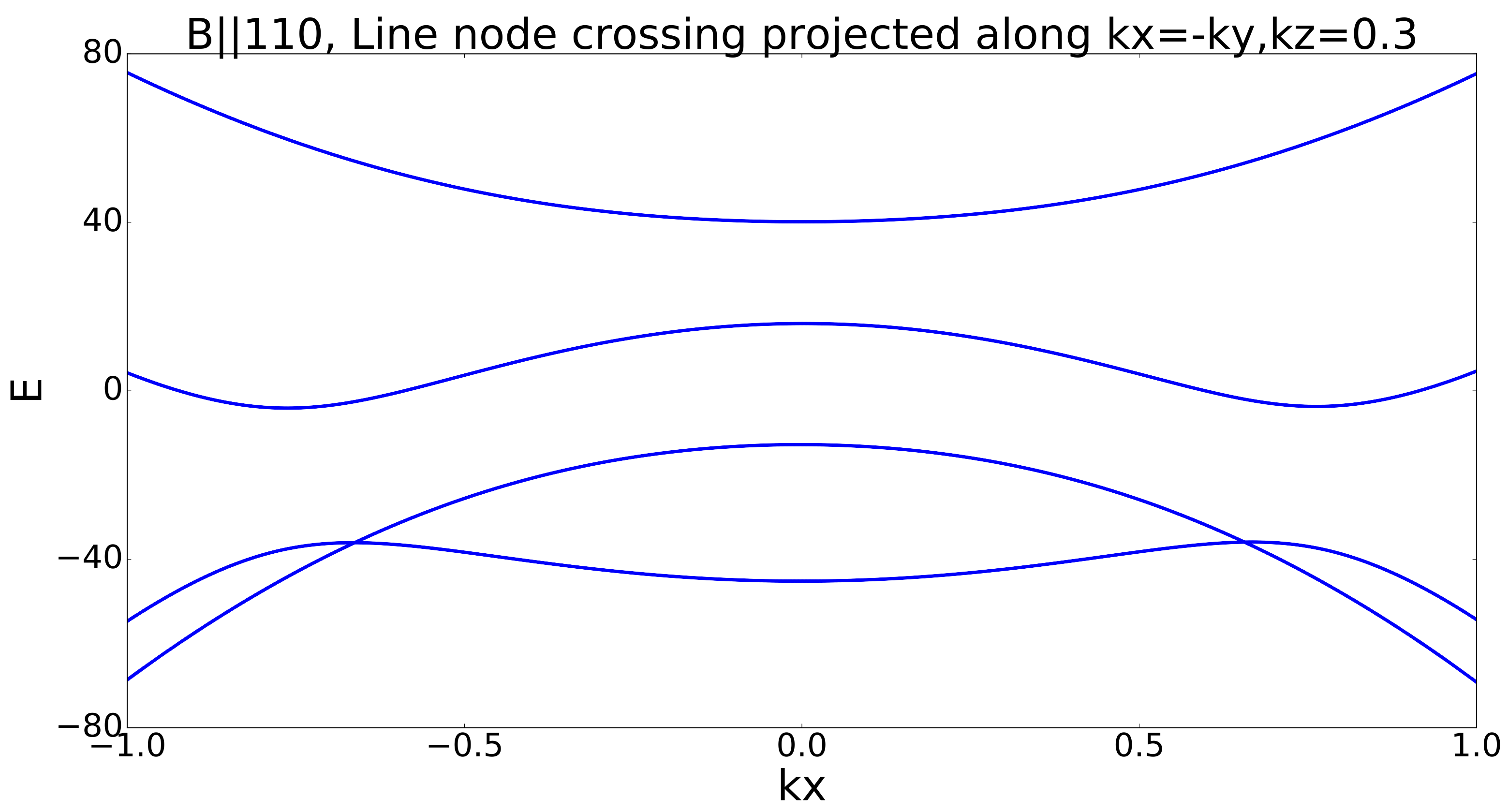}}
\subfloat[\label{fig:111}]{
\includegraphics[width=0.4\textwidth]{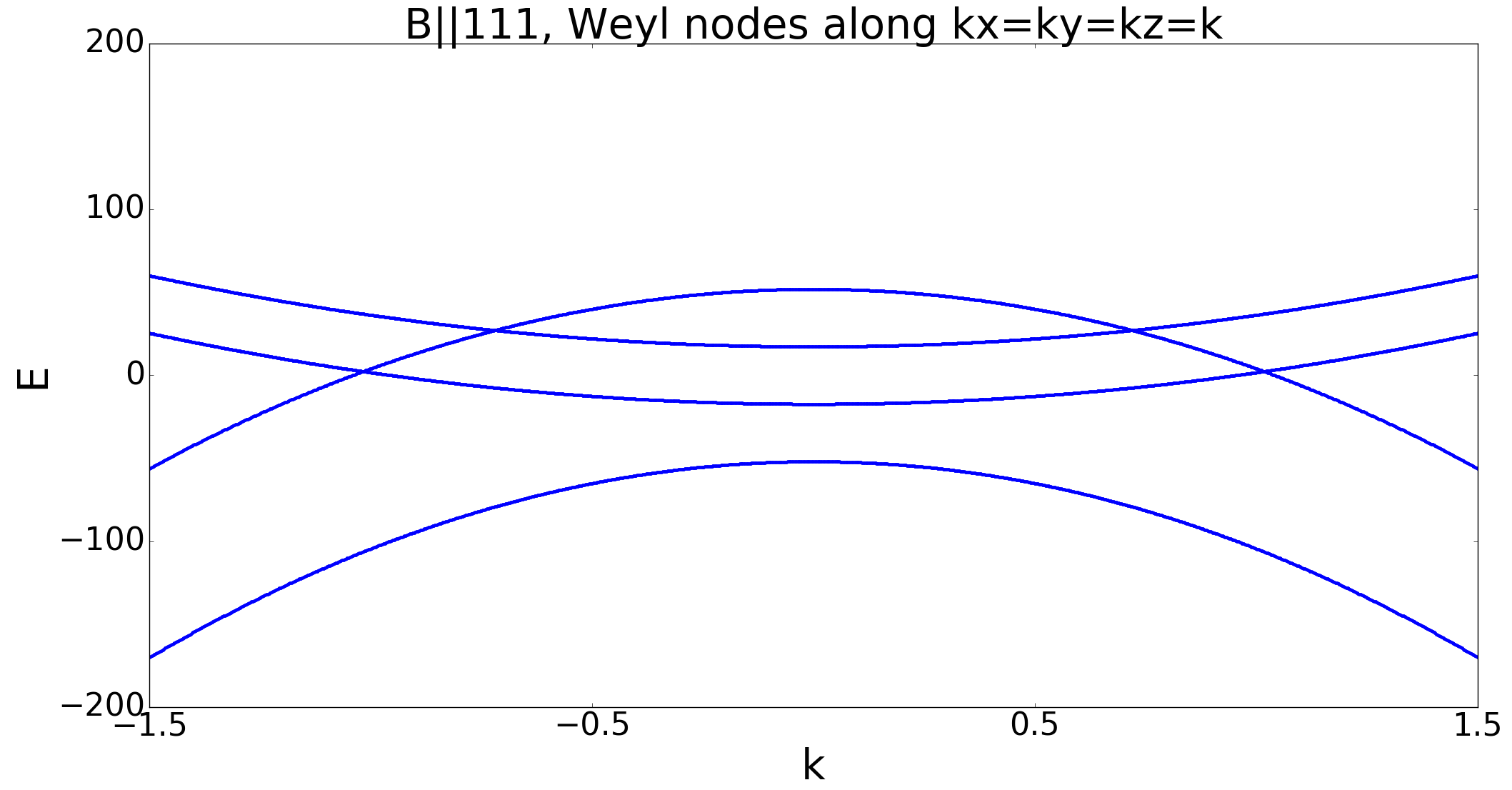}}
\caption{(a) shows the spectrum with $\vec{B}=(20,20,0)$, plotted along the high symmetry line $\vec{k}=(k_x,-k_x,0)$. The two crossings extend to line nodes for $k_z\neq 0$. This can be seen from the linearized effective Hamiltonian, which near the left-most node has $\vec{v}=(-17.84,17.84,0)$ and $A=\left(
\begin{array}{ccc}
0 & 43.56 & 17.65 \\
0 & 43.56 & 17.65 \\
0 & 0 & 0
\end{array}
\right)
$ (b) Shows the spectrum with $\vec{B}=(20,20,20)$, plotted along $\vec{k}=(k,k,k)$. There are four Weyl nodes protected by $C_{3,111}$ symmetry. At $B=20(1,1,1)$, we have computed the velocities and $A$ matrix to verify that the crossings are linear and that the Weyls are Type I. For the nodes at $k\approx .707$ and $k\approx 1.00$ on the plot, we find the velocities $v=-6.938(1,1,1)$ and $v=-9.812(1,1,1)$, respectively, with corresponding matrices:
$
A=\left(
\begin{array}{ccc}
 24.15 & -39.93 & -16.33 \\
 22.51 & 40.88 & -16.33 \\
 -46.66& -0.95& -16.33 \\
\end{array}
\right)
 \text{ and } A=\left(
\begin{array}{ccc}
 -32.02 & 58.48 & -23.09 \\
 -34.63 & 56.97 & -23.09 \\
 66.65& 1.510& -23.09 \\
\end{array}
\right)
$\\
The $C_{3,111}$ symmetry dictates that the velocities are the same and that the entries in the last column of the $A$ matrix are the same. We have checked that $(AA^T)_{ij}-v_iv_j$ has three positive eigenvalues in both cases.}
\end{figure*}

%\section{C\MakeLowercase{d}$_3$A\MakeLowercase{s}$_2$}
\section{C\lowercase{d}$_3$A\lowercase{s}$_2$}
\subsection{$k\cdot p$ Hamiltonian}

In Cd$_3$As$_2$, there are eight bands of interest: two $s$-orbitals with $|J,J_z\rangle=|1/2,\pm 1/2\rangle$ and six $p$-orbitals with $|J,J_z\rangle=|3/2,\pm 3/2\rangle, |3/2,\pm 1/2\rangle, |1/2, \pm 1/2\rangle$. In the presence of spin-orbit coupling, the states nearest to the Fermi level come from the heavy-hole $p$-states $|3/2,\pm 3/2\rangle$ and the conduction $s$-states $|1/2,\pm 1/2\rangle$\cite{WangWengWuEtAl2013}. Since the bands are inverted (the $p$-orbitals disperse downwards and the $s$-orbitals upwards), they will intersect at some point in the Brillouin zone, but not at the $\Gamma$ point. These facts comprise the major difference between the previous analysis of GdPtBi and the current analysis of Cd$_3$As$_2$: in the former case, the low-energy bands comprise a four-dimensional irreducible representation (irrep) and, consequently, form a four-band multiplet at the $\Gamma$ point; in contrast, the low-energy bands in C$_3$As$_2$ come from two two-dimensional irreps, which are not generically degenerate at the $\Gamma$ point, but must cross somewhere in the Brillouin zone. In both cases, band inversion is crucial for Weyl points to emerge in the presence of a magnetic field.

We work in the basis $j_z=3/2, 1/2, -1/2, -3/2$. 
We take the $D_{4h}$ symmetry group for Cd$_3$As$_2$\cite{Ali14}, which has generators
\begin{equation} C_{4z} = \begin{pmatrix}
 -(-1)^{1/4} & 0 & 0 & 0 \\
 0 & -(-1)^{3/4} & 0 & 0 \\
 0 & 0 & (-1)^{1/4} & 0 \\
 0 & 0 & 0 & (-1)^{3/4} \end{pmatrix}
 \label{eq:C4z}
\end{equation}
\begin{equation} C_{2x} = \begin{pmatrix}
 0 & 0 & 0 & i \\
 0 & 0 & -i & 0 \\
 0 & -i & 0 & 0 \\
 i & 0 & 0 & 0 \end{pmatrix} \end{equation}
\begin{equation} M_{001} = \begin{pmatrix} 
 -i & 0 & 0 & 0 \\
 0 & -i & 0 & 0 \\
 0 & 0 & i & 0 \\
 0 & 0 & 0 & i \end{pmatrix}\end{equation}
in addition to time reversal, which takes the form
\begin{equation} T = \begin{pmatrix}
 0 & 0 & 0 & -1 \\
 0 & 0 & -1 & 0 \\
 0 & 1 & 0 & 0 \\
 1 & 0 & 0 & 0 \end{pmatrix} K \end{equation}
These symmetries imply $C_{2z}$, $C_{2y}$, $C_{2,110}$, $C_{2,1\bar{1}0}$, $M_{100}$, $M_{010}$, $M_{110}$, $M_{1\bar{1}0}$, inversion and the product of inversion and $C_{4z}$.

The $k\cdot p$ Hamiltonian to order $k^3$ derived from the symmetries above takes the form:
\begin{equation}
H_0(k) = \epsilon_0(k)+
\begin{pmatrix}
\mathcal{M}(k) & \mathcal{A}(k) & \mathcal{B}(k) & 0 \\
 \mathcal{A}^*(k) & -\mathcal{M}(k) & 0 & \mathcal{B}(k) \\
 \mathcal{B}^*(k) & 0 & -\mathcal{M}(k) & - \mathcal{A}(k) \\
 0 & \mathcal{B}^*(k) & - \mathcal{A}^*(k) &\mathcal{M}(k) \\
\end{pmatrix}
\label{eq:HCdAs}
\end{equation}
where 
\begin{align}
\mathcal{M}(k) &\equiv  M_0 + M_1 k_z^2+ M_2k_\parallel^2\nonumber\\
\mathcal{A}(k) &\equiv Ak_- \left( 1 + A_z k_z^2 \right) + k_+\left( A_1k_+^2 +A_2k_-^2  \right) \nonumber\\
\mathcal{B}(k) &\equiv \left(B_1k_+^2 + B_2 k_-^2 \right)k_z\nonumber\\
\epsilon_0(k) &\equiv C_0 + C_1 k_z^2+ C_2k_\parallel^2
\end{align}
and $k_{\pm} = k_x \pm i k_y$ and $k_\parallel^2 = k_x^2 + k_y^2=k_+k_-$.

The energies of this Hamiltonian can be solved for exactly and are found to be doubly degenerate everywhere:
\begin{equation} E(k) = \epsilon_0(k) \pm \sqrt{\mathcal{M}(k)^2 +| \mathcal{A}(k)|^2 + |\mathcal{B}(k)|^2 }\end{equation}

It is evident that the doubly-degenerate bands cross each other when $\mathcal{M}(k) = \mathcal{A}(k) = \mathcal{B}(k) = 0$. For generic parameters, this occurs when $k_z=\pm\sqrt{-M_0/M_1},k_\parallel=0$.
Thus, it is crucial that $M_0$ and $M_1$ have opposite signs, in agreement with the fit in Eq~(\ref{eq:abinitioparamsCdAs}).
%(One might wonder whether there is a second band crossing when $k_z=0, k_\parallel=\pm \sqrt{-M_0/M_2}$, which implies $\mathcal{B}(k)=0$, but then $\mathcal{A}(k)=0$ generates two additional equations for $k_x, k_y$ -- one from its real part and one from its imaginary part -- which is generically too constraining.) 

\subsection{Material parameters}

%To order $k^2$, the Hamiltonian (\ref{eq:HCdAs}) is block diagonal ($\mathcal{B}(k)$ appears only at order $k^3$) and doubly degenerate, with eigenvalues $E(k) =  \epsilon_0(k) \pm \sqrt{M(k)^2 + A^2 k_\parallel^2}$.
%\begin{equation}
%H_0(k) = \epsilon_0(k)+
%\begin{pmatrix}
%\mathcal{M}(k) & Ak_- & 0 & 0 \\
 %Ak_+ & -\mathcal{M}(k) & 0 & 0 \\
 %0 & 0 & -\mathcal{M}(k) & -Ak_- \\
 %0 & 0 & -Ak_+ & \mathcal{M}(k) \\
%\end{pmatrix} + O(k^3),
%\end{equation}
Ab initio calculations give the order $k^2$ parameters in (\ref{eq:HCdAs}):
\begin{equation}
\begin{array}{cccccccc}
M_0 &= .0205, & M_1 &= -18.77, & M_2 &= -13.5, & A &= 0.889\\
C_0 &= -.0145, & C_1 &= 10.59, &C_2 &= 11.5 & &\\
\end{array}
\label{eq:abinitioparamsCdAs}
\end{equation}
Energy is in units of eV when momentum is in units \AA$^{-1}$.

%%%%%%%%%%%%%%%%%%%%%%%
%%%%%%%%%%%%%%%%%%%%%%%

\subsection{Magnetic field}

The Zeeman term takes the form:
\begin{equation} H_Z = -(g_s J_s+ g_p J_p) \cdot B \label{eq:ZeemanCd3As2}\end{equation}
where, in our basis, 
\begin{equation} (J_s)_{i} = \begin{pmatrix} 0 & \begin{matrix}0 & 0\end{matrix} &0 \\ 0 & \left(\frac{1}{2}\sigma_i\right) & 0 \\ 0 &  \begin{matrix}0 & 0\end{matrix} &0\end{pmatrix}, (J_p)_z=\begin{pmatrix} \frac{3}{2} & 0 & 0 & 0 \\ 0 & 0 & 0 & 0 \\0 & 0 & 0 & 0 \\ 0 & 0 & 0 & -\frac{3}{2}\end{pmatrix}
\label{eq:JsJp} \end{equation}
Since $(J_p)_{x,y}$ mix with bands further away in energy that are not included in the $k\cdot p$, we determine their effective forms perturbatively below when necessary.
Notice that in (\ref{eq:ZeemanCd3As2}), we have included two $g$-factors because the bands come from different representations.

We now consider the band crossings when the magnetic field is along one of the high-symmetry axes.

\subsection{Field in the $\hat{z}$ direction}

The $k_z$-axis is invariant under $C_{4z}$. Since $H_0+ H_Z$ is diagonal along this axis when $B=B_z\hat{z}$, Eq~(\ref{eq:C4z}) shows that each of the four bands has a different eigenvalue of $C_{4z}$. Hence, band crossings along this axis are protected; we show below that there are between four and eight band crossings along this axis, depending on the magnitude of $B_z$ and $g_{p,s}$.

The $k_z=0$ plane is also a high-symmetry plane, invariant under $M_{001}= {\rm Diag}[-i,-i,i,i]$. For small $B_z$, this plane is gapped. As $B_z$ is increased, line nodes emerge between different bands depending on the relative signs and magnitudes of $g_s$ and $g_p$.

We now compute the effective two-band Hamiltonians that describe the band crossings along the $k_z$ axis.
%; we utilize the order $k^2$ Hamiltonian,
%\begin{widetext}
%\begin{equation}
%H(k)\Big|_{B=B_z\hat{z}} = \epsilon_0(k)+
%\begin{pmatrix}
%\mathcal{M}(k)- \frac{3}{2}g_pB_z & \mathcal{A}(k) & \mathcal{B}(k) & 0 \\
 %\mathcal{A}^*(k) & -\mathcal{M}(k)-\frac{1}{2}g_sB_z & 0 & \mathcal{B}(k) \\
 %\mathcal{B}^*(k) & 0 & -\mathcal{M}(k)+\frac{1}{2}g_sB_z & - \mathcal{A}(k) \\
 %0 & \mathcal{B}^*(k) & - \mathcal{A}^*(k) &\mathcal{M}(k)+\frac{3}{2}g_pB_z \\
%\end{pmatrix}
%\end{equation}
%\end{widetext}
We first consider the crossing between the $j_z=3/2$ and $j_z=1/2$ bands, which occurs at $k_{z0}^2 = ((g_s-3g_p)B_z/4-M_0)/M_1$. The linearized Hamiltonian is:
\begin{equation} H_{\rm eff}^{\frac{3}{2}\frac{1}{2}}=2M_1k_{z0}\delta_{z}\sigma_z + a(k_x\sigma_x + k_y\sigma_y),
\end{equation}
%\begin{equation}\begin{pmatrix}2M_1k_{z0}\delta_{k_z} & a\left(\delta_{k_x}-i\delta_{k_y}\right) \\ a\left(\delta_{k_x}-i\delta_{k_y}\right) & -2M_1k_{z0}\delta_{k_z} \end{pmatrix}\end{equation}
where $a=A(1+A_zk_{z0}^2)$ and $\delta_z \equiv k_z-k_{z0}$.
%Using our usual notation, where a Weyl point is described by a vector $v$ and a matrix $\bar{A}$, this describes a Weyl point with parameters $v=0$ and $\bar{A}_{xx} = \bar{A}_{yy} = a, \bar{A}_{zz} = 2M_1k_{z0}$. Hence, $\det\bar{A} = 2M_1A^2(1+A_zk_{z0}^2)k_{z0} \sim |g_2B_z - M_0|^{1/2}(1+A_z(g_2B_z-M_0))$.
%In the small $B$ limit, $\det \bar{A}\sim 1+B_z^{1/2}$.
This band crossing describes a single Weyl point.
The band crossing between $j_z = -3/2$ and $j_z = -1/2$ is similar: the Hamiltonian has an overall negative sign and $k_{z0}$ is the same after the replacement $B_z\rightarrow -B_z$. For any magnitude of $|B|$, at least one of these pairs of single Weyl points will exist; for small enough $|B|$, both pairs exist.

The band crossing between the $j_z = 3/2$ and $-1/2$ bands occurs at $k_{z0}^2 = \left(  -B_z(g_s + 3g_p)/4- M_0\right)/M_1$. The effective Hamiltonian is:
\begin{equation} H_{\rm eff}^{\frac{3}{2},-\frac{1}{2}}= 
\begin{pmatrix} 2M_1k_{z0}\delta_{z} & 2\left(B_2 k_-^2+B_1k_+^2\right)k_{z0}\\
2\left(B_2 k_+^2+B_1k_-^2\right)k_{z0} & -2M_1k_{z0}\delta_{z} \end{pmatrix},
\end{equation}
which describes a double Weyl point.
The crossing between the $j_z = -3/2$ and $1/2$ bands is similar: the diagonal entries of the Hamiltonian have the opposite sign, while the off-diagonal entries are the same, and $k_{z0}$ is found by taking $B_z\rightarrow -B_z$.
For any magnitude of $|B|$, at least one of these pairs of double Weyl points will exist; for small enough $|B|$, both pairs exist.

Thus, when $B=B_z\hat{z}$, at least one pair of single and one pair of double Weyl points emerge on the $k_z$ axis. For small enough $B$, two pairs of single and two pairs of double Weyl points emerge.

\subsection{Field in the $\hat{x}$ direction}

The $k_x$ axis is invariant under $C_{2x}$ symmetry. This can protect crossings between any two bands with opposite $C_{2x}$ eigenvalues, for a maximum of four total Weyl points.

The $k_x=0$ plane is invariant under the mirror symmetry, $M_{100}$. Under $M_{100}$, one of the $s$-orbitals has eigenvalues $+i$ and the other $-i$; the same holds for the $p$-orbitals. This protects up to two line nodes, between the $s$-orbital with $j_x=\pm 1/2$ and the $p$-orbital with $j_x=\mp3/2$. For any value of $B_x$, it is guaranteed that at least one of these will exist; if $|B|$ is small enough, both will appear. The other two potential crossings are avoided.

The $k_y=0$ plane is also a high-symmetry plane, invariant under the product $C_{2y}T$, but 
%However, this symmetry does not protect line nodes (we can see this by choosing the $T = \sigma_y K$, $C_{2y} = i\sigma_y$, which, when imposed as a symmetry on an effective Hamiltonian $d_i(k_x,k_z)\sigma_i$ only forces $d_y=0$, leaving two parameters and two variables, which does not protect a point node.) 
this plane is generically gapped.

We now now verify these claims using the $k\cdot p$ model: only two of the four $p$-orbitals with $J=3/2$ are included in the low-energy $k\cdot p$ model (\ref{eq:HCdAs}); in particular, the $J_z=\pm 3/2$ orbitals are included while the $J_z=\pm 1/2$ orbitals are not. However, the Zeeman term will mix all four orbitals. Assume the $J_z=\pm 1/2$ orbitals are separated by some energy, $\Delta$, from the $J_z=\pm 3/2$ bands in the $k\cdot p$ model (\ref{eq:HCdAs}).
Then when $B=B_x\hat{x}$, the Zeeman term can be added perturbatively to (\ref{eq:HCdAs}), which yields the following Hamiltonian,
%\begin{widetext}
\begin{equation}
H(k)\! \!=\! \epsilon_0(k)+\!
\begin{pmatrix}
\!\mathcal{M}(k)\!+\!B_x\delta\! & \mathcal{A}(k) & \mathcal{B}(k) & -B_x\delta^2 \\
 \mathcal{A}^*(k) & -\mathcal{M}(k) & \!-\frac{1}{2}g_sB_x\! & \mathcal{B}(k) \\
 \mathcal{B}^*(k) & \!-\frac{1}{2}g_sB_x\! & -\mathcal{M}(k) & -\mathcal{A}(k) \\
 -B_x\delta^2 & \mathcal{B}^*(k) & -\mathcal{A}^*(k) & \!\mathcal{M}(k)\!+\!B_x\delta\! \\
\end{pmatrix}
\end{equation}
%\end{widetext}
where $\delta \propto B_x/\Delta$. This is the lowest order term that is consistent with $C_{2x}$ symmetry, which remains a symmetry when $B=B_x\hat{x}$.

Since the $k_x$ axis is gapped when $B_x=0$, Weyl points do not emerge along this axis until $B_x$ is large enough. When this value is reached, two or four Weyl points emerge. By computing effective two-band Hamiltonians, we have verified that these are all single Weyl points.

%We would also like to investigate the $k_z$-axis; here the second term in the Hamiltonian simplifies to:
%\begin{equation}
%\begin{pmatrix}
%\mathcal{M}(k)+B_x\delta & 0 & 0 & -B_x\delta^2 \\
%0 & -\mathcal{M}(k) & -\frac{1}{2}g_sB_x & 0 \\
 %0 & -\frac{1}{2}g_sB_x & -\mathcal{M}(k) & 0 \\
 %-B_x\delta^2 & 0 & 0 & \mathcal{M}(k)+B_x\delta \\
%\end{pmatrix}
%\end{equation}
Along the $k_z$ axis, the eigenvalues are given by,
\begin{align}
E_{+\pm} &=\mathcal{M}(k)+B_x\delta \pm B_x\delta^2 \nonumber\\
E_{-\pm} &= -\mathcal{M}(k) \pm \frac{1}{2}g_sB_x
\end{align}
There are four possible crossings, either $E_{+\pm}=E_{-\pm}$ or $E_{+\pm} = E_{-\mp}$.

In the first case, the band crossings occur at $k_{z0}$ satisfying $2\mathcal{M}(k_{z0})=B_x(\frac{g_s}{2}-\delta - \delta^2)$.
To leading order, the effective Hamiltonian between the bands $E_{++}$ and $E_{-+}$ is given by,
\begin{eqnarray}
&H_{\rm eff} = -2\left(\delta_{z}k_{z0}M_1 +(k_x^2+k_y^2)M_2\right)\sigma_z+ \nonumber\\
&-Ak_x\left( 1 + A_zk_{z0}^2\right) \sigma_x + 4k_{z0}(B_2-B_1)k_xk_y\sigma_y
\end{eqnarray}
where $\delta_z\equiv k_z-k_{z0}$.
This is part of the line node in the $k_x=0$ plane: setting $k_x=0$, $H_{\rm eff} \rightarrow -2(\delta_{z}k_{z0}M_1 + k_y^2M_2)\sigma_z$, for which there is a line of degenerate eigenvalues.
The effective Hamiltonian between $E_{+-}$ and $E_{--}$ is the complex conjugate of this one.

The other band crossings occur at $k_{z1}$ satisfying $2\mathcal{M}(k_{z1}) = B_x(-\frac{g_s}{2}-\delta +\delta^2)$.
The effective Hamiltonian at the crossing between $E_{++}$ and $E_{--}$ is, to leading order,
\begin{eqnarray}
&H_{\rm eff} = -2\left(\delta_{z}k_{z1}M_1 + (k_x^2+k_y^2)M_2\right)\sigma_z\label{eq:nonWeyl}\\
&-2(B_1+B_2)(k_x^2-k_y^2)k_{z1}\sigma_x -Ak_y\left(1+A_zk_{z1}^2\right)\sigma_y\nonumber
\end{eqnarray}
This describes a point crossing, however, the Chern number is zero. Thus, it is not a Weyl point. 
%We can also see this by adding an arbitrarily small perturbation $\alpha(-2(B_1+B_2)k_{z1})\sigma_x$ for some $\alpha>0$: a Weyl point would still have a band crossing after adding a small perturbation; however, here we see that a band crossing requires $\delta_{k_y}=0, \delta_{k_x}^2-\delta_{k_y}^2+\alpha=0$, for which there is no solution. Hence, the crossing described by (\ref{eq:nonWeyl}) is an accidental degeneracy. The same is true for the crossing between $E_{+-}$ and $E_{-+}$, which has the same Hamiltonian with the opposite coefficient of $\sigma_x$.
%At $B=0$, this band crossing is ``protected'' by symmetries of (\ref{eq:HCdAs}) that are no longer symmetries when we turn on the magnetic field $B=B_x\hat{x}$. For example, the crossing would be avoided by a term proportional to some power of $k_z$ added to $\mathcal{B}(k)$, which mixes the $\pm 3/2$ states with the $\pm 1/2$ states; this is forbidden by $C_{4z}, M_{110}, M_{1\bar{1}0}, C_{2,110}$, and $C_{2,1\bar{1}0}$, but none of these symmetries survive when we turn on $B_x\hat{x}$.

The same analysis holds when $B$ is in the $[010],[110]$, or $[1\bar{1}0]$ directions, which also have $C_2$ symmetry.

\subsection{Band crossings along the $k_z$ axis for arbitrary $\mathbf{B}$}

On the $k_z$ axis, symmetry constrains the $k\cdot p$ Hamiltonian to the form: $H(k_z) = \epsilon'(k_z) + \mathcal{M}'(k_z){\rm diag}[1,-1,-1,1]$, where $\epsilon'(k_z)$ and $\mathcal{M}'(k_z)$ are even functions of $k_z$ that are equal to $\epsilon_0(k_z)$ and $\mathcal{M}(k_z)$ to order $k_z^2$.
Thus, the Hamiltonian with the Zeeman term, $H_Z$ in Eq~(\ref{eq:ZeemanCd3As2}), splits into two $2\times 2$ blocks, $\epsilon'(k_z)+\mathcal{M}'(k_z) -g_pJ_p\cdot B$ and $\epsilon'(k_z)-\mathcal{M}'(k_z)-g_sJ_s\cdot B$.
The eigenvalues of $g_sJ_s\cdot B$ are $\pm \frac{g_s}{4}|B|$.
From the previous subsection, we know the eigenvalues of $g_pJ_p\cdot B$ do not take such a simple form, but take the form $E_0(B_{x,y})\pm E_{p}(B_{x,y,z})$. 
Since $H_Z$ has no zero eigenvalues for finite $B$, there will be band crossings along the $k_z$ axis when $2\mathcal{M}'(k_z) = E_0\pm E_p \mp \frac{g_s}{4}|B|$, where the $\pm$ signs are uncorrelated.
As long as $E_0 < M_0$ (recall from the previous subsection that $E_0\sim 1/\Delta$, where $\Delta$ is the splitting between the $p$ orbitals with $|J_z|= 1/2$ and $J_z=|3/2|$), this equation will always have solutions.
However, as we saw in the previous subsection, these band crossings are not always Weyl points.

%%%%%%%%%%%%%%%%%%%%%%%
%%%%%%%%%%%%%%%%%%%%%%%

%\subsection{Summary}
%
%When $B$ is in the $\hat{z}$ direction:
%\begin{itemize}
%\item There are between four and eight Weyl points on the $k_z$ axis (depending on the magnitudes of $B_z$ and the $g$-factors), protected by $C_{4z}$
%\item The crossings between $j_z=\pm 3/2$ and $j_z=\pm 1/2$ (the same $\pm$ sign) are single Weyl points, with $\det A \sim 1+B_z^{1/2}$ in the small $B$ limit
%\item The other four crossings are double Weyl points.
%\item There are line nodes in the $k_z=0$ plane protected by $M_{001}$ (it only protects half of the potential band crossings)
%\end{itemize}
%
%When $B$ is in the $\hat{x}$ direction (same analysis applies to $\hat{y}$ and $\hat{x}\pm \hat{y}$)
%\begin{itemize}
%\item There are up to four Weyl points on the $k_x$ axis, protected by $C_{2x}$
%\item There are line nodes in the $k_x = 0$ plane protected by $M_{100}$, which protects half of the potential band crossings
%\end{itemize}
%
%It would be good to check these results numerically to confirm that we did not miss any high-symmetry directions and to confirm our understanding of nodes vs points.
%
%As explained in the main text, we also saw accidental degeneracies, where apparent band crossings that result from the $k\cdot p$ can be avoided if terms are added that are consistent with the symmetries in the presence of a magnetic field, but not consistent with the full crystal symmetries.
%\section{N\MakeLowercase{a}$_3$B\MakeLowercase{i}}
\section{N\lowercase{a}$_3$B\lowercase{i}}
\subsection{$k\cdot p$ Hamiltonian}

The analysis proceeds similar to Cd$_3$As$_2$: in Na$_3$Bi there are eight bands of interest, two $s$-orbitals with $(J,J_z)=(1/2,\pm 1/2)$ and six $p$-orbitals with $(J,J_z)=(3/2,\pm 3/2), (3/2,\pm 1/2), (1/2, \pm 1/2)$. Following Ref~[\onlinecite{WangSunChenEtAl2012}], in the presence of spin-orbit coupling, the states nearest to the Fermi level come from the heavy-hole $p$-states $|j,j_z\rangle = |3/2,\pm 3/2\rangle $ and the conduction $s$-states $|1/2,\pm 1/2\rangle$. Thus, the system is represented by two two-dimensional irreps. Since the bands are inverted (the $p$-orbitals disperse downwards and the $s$-orbitals upwards), they will intersect at some point in the Brillouin zone away from the $\Gamma$ point.
Below, we work in the basis ordered by $j_z=3/2, 1/2, -1/2,-3/2$.

%%%%%
%%%%% OLD BASIS: jz = 3/2, -3/2, 1/2, -1/2
%%%%%
%We take the $D_{6h}$ symmetry group for NaBi, which has generators
%\begin{equation} C_{6z} = \begin{pmatrix}
 %-i & 0 & 0 & 0 \\
 %0 & i & 0 & 0 \\
 %0 & 0 & -(-1)^{5/6} & 0 \\
 %0 & 0 & 0 & (-1)^{5/6}
%\end{pmatrix}\label{eq:C6z}
%\end{equation}
%\begin{equation} C_{2x} = \begin{pmatrix}
 %0 & i & 0 & 0 \\
 %i & 0 & 0 & 0 \\
 %0 & 0 & 0 & -i \\
 %0 & 0 & -i & 0 
%\end{pmatrix} \end{equation}
%\begin{equation} M_{100} = \begin{pmatrix} 
 %0 & -i & 0 & 0 \\
 %-i & 0 & 0 & 0 \\
 %0 & 0 & 0 & -i \\
 %0 & 0 & -i & 0 \end{pmatrix}\end{equation}
%in addition to time reversal, which takes the form
%\begin{equation} T = \begin{pmatrix}
 %0 & -1 & 0 & 0 \\
 %1 & 0 & 0 & 0 \\
 %0 & 0 & 0 & -1 \\
 %0 & 0 & 1 & 0 \end{pmatrix} K \end{equation}
We take the $D_{6h}$ symmetry group for NaBi, which has generators
\begin{equation} C_{6z} = \begin{pmatrix}
 -i & 0 & 0 & 0 \\
 0 & -(-1)^{5/6} & 0 & 0 \\
 0 & 0 & (-1)^{5/6} & 0 \\
 0 & 0 & 0 & i
\end{pmatrix}\label{eq:C6z}
\end{equation}
\begin{equation} C_{2x} = \begin{pmatrix}
 0 & 0 & 0 & i \\
 0 & 0 & -i & 0 \\
 0 & -i & 0 & 0 \\
 i & 0 & 0 & 0 
\end{pmatrix} \end{equation}
\begin{equation} M_{100} = \begin{pmatrix} 
 0 & 0 & 0 & -i \\
 0 & 0 & -i & 0 \\
 0 & -i & 0 & 0 \\
 -i & 0 & 0 & 0 \end{pmatrix}\end{equation}
in addition to time reversal, which takes the form
\begin{equation} T = \begin{pmatrix}
 0 & 0 & 0 & -1 \\
 0 & 0 & -1 & 0 \\
 0 & 1 & 0 & 0 \\
 1 & 0 & 0 & 0 \end{pmatrix} K \end{equation}

These symmetries imply $C_{3z}$; $C_2$ rotations about and mirror planes perpendicular to: $\hat{y},\hat{z}, \hat{a}, \hat{b},\hat{c},\hat{d}$; and inversion. We have defined the unit vectors corresponding to $\pi/6$ rotations of $\hat{x}$:
\begin{align}
\hat{a} &= \frac{1}{2}\left( \hat{x} + \sqrt{3}\hat{y}\right)\nonumber\\
\hat{b} &= \frac{1}{2}\left( -\hat{x} + \sqrt{3}\hat{y}\right)\nonumber\\
\hat{c} &= \frac{1}{2}\left( \sqrt{3}\hat{x} +\hat{y}\right)\nonumber\\
\hat{d} &= \frac{1}{2}\left( -\sqrt{3}\hat{x} +\hat{y}\right)
\label{eq:symmvecs}
\end{align}

The $k\cdot p$ Hamiltonian to order $k^3$ derived from the symmetries above has the same structure as that of Cd$_3$As in Eq~(\ref{eq:HCdAs}), but with the matrix elements,
%\begin{equation}
%H_0(k) = \epsilon_0(k)+
%\begin{pmatrix}
%\mathcal{M}(k) & 0 & \mathcal{A}(k) & \mathcal{B}(k) \\
%0 & \mathcal{M}(k) & \mathcal{B}^*(k) & -\mathcal{A}^*(k)\\
 %\mathcal{A}^*(k) & \mathcal{B}(k) & -\mathcal{M}(k) & 0 \\
 %\mathcal{B}^*(k) & -\mathcal{A}(k) & 0 &-\mathcal{M}(k) \\
%\end{pmatrix}
%\label{eq:HNaBi}
%\end{equation}
%where 
\begin{align}
\mathcal{M}(k) &\equiv  M_0 + M_1 k_z^2+ M_2k_\parallel^2\nonumber\\
\mathcal{A}(k) &\equiv Ak_- \left( 1 + A_1 k_z^2+A_2k_\parallel^2 \right)  \nonumber\\
\mathcal{B}(k) &\equiv -\frac{1}{2}Bk_-^2k_z\nonumber\\
\epsilon_0(k) &\equiv C_0 + C_1 k_z^2+ C_2k_\parallel^2,
\end{align}
where $k_{\pm} = k_x \pm i k_y$ and $k_\parallel^2 = k_x^2 + k_y^2=k_+k_-$.

The energies of this Hamiltonian can be solved for exactly and are doubly degenerate everywhere:
\begin{equation} E(k) = \epsilon_0(k) \pm \sqrt{\mathcal{M}(k)^2 +| \mathcal{A}(k)|^2 + |\mathcal{B}(k)|^2 }\end{equation}
The doubly-degenerate bands cross each other when $\mathcal{M}(k) = \mathcal{A}(k) = \mathcal{B}(k) = 0$. This occurs when $k_z=\pm\sqrt{-M_0/M_1},k_\parallel=0$.
Thus, it is crucial that $M_0$ and $M_1$ have opposite signs, as agrees with the fit in Eq~(\ref{eq:params}).

\subsection{Material parameters}

Ref~[\onlinecite{WangSunChenEtAl2012}] provides the fit to the coefficients of the quadratic terms:
\begin{align}
M_0 &= 0.08686\nonumber\\
M_1 &= -10.6424\nonumber\\
M_2 &= -10.361\nonumber\\
C_0 &= -0.06382\nonumber\\
C_1 &= 8.7536\nonumber\\
C_2 &= -8.4008\nonumber\\
A &= 2.4598 \label{eq:params}
\end{align}
where the energy is in units of eV when the momentum is in units \AA$^{-1}$.

\subsection{Magnetic field}

The Zeeman term takes the form:
\begin{equation} H_Z = -(g_s J_s+ g_p J_p) \cdot B \label{eq:Zeeman}\end{equation}
where, if we exclude mixing with bands outside the $k\cdot p$ model, the $J_s$ and $J_p$ matrices are given by Eq~(\ref{eq:JsJp}). 
%\begin{equation} (J_s)_{i} = \begin{pmatrix}0 & 0 \\ 0 & \sigma_i\end{pmatrix}, (J_p)_z=\begin{pmatrix} \frac{3}{2}\sigma_z & 0 \\ 0 & 0\end{pmatrix} 
%\end{equation}
As in the previous section, the $p$-orbitals with $J=3/2$, $J_z=\pm 1/2$ are not included in the $k\cdot p$. 
We will include the mixing between these orbitals and the $J_z=\pm 3/2$ orbitals perturbatively below when needed.

\subsection{Field in the $\hat{z}$ direction}

When the magnetic field is in the $\hat{z}$ direction, the $k_z$-axis remains a high-symmetry line and $C_{6z}$ preserves points on this line. From Eq~(\ref{eq:C6z}), it is evident that all bands have different eigenvalues under $C_{6z}$; hence crossings between all bands are protected. As in Cd$_3$As$_2$, this can yield up to eight Weyl points (and a minimum of four.)

The crossing between $j_z=\pm 1/2$ and $j_z = \mp 3/2$ occurs at $(0,0,k_{z0})$, where $k_{z0}^2 =  \left( \pm B_z(g_s + 3g_p)/4 - M_0\right)/M_1$; the crossing point can be easily found, since $H_0(0,0,k_z)$, $(J_s)_z$ and $(J_p)_z$ are all diagonal. The effective two-band Hamiltonian describing the band crossing is given by,
\begin{eqnarray}
&H_{\rm eff}^{\pm \frac{1}{2},\mp\frac{3}{2}} =
%-2(M_1\delta_{z}^2+M_1k_{z0}\delta_{z} + (k_x^2+k_y^2)M_2)\sigma_z 
-2(M_1k_{z0}\delta_{z} + (k_x^2+k_y^2)M_2)\sigma_z \\
&-B(k_x^2 -k_y^2)k_{z0}\sigma_x \pm 2Bk_xk_yk_{z0}\sigma_y,\nonumber
\end{eqnarray}
which describes a double Weyl point at $k_{z0}$.

The other crossings are single Weyl points, described by the two-band Hamiltonian: $H_{\rm eff}^{\pm \frac{1}{2},\pm\frac{3}{2}}=-2\delta_{z}k_{z0}M_1\sigma_z +A(\pm k_x\sigma_x + k_y\sigma_y)$, where here $k_{z0}^2 =  \left( \pm B_z(g_s - 3g_p)/4 - M_0\right)/M_1$. 
%The determinant of the ``$A$'' matrix is $-2M_1A^2k_{z0} \sim (1+B_z^{1/2})$.

The $k_z=0$ plane is a high-symmetry plane protected by $M_{001}$; since mirror symmetries have two distinct eigenvalues, this protects two of the four possible crossings. These protected crossings yield line nodes.

\subsection{Field in the $\hat{x}$ direction}

When the magnetic field is in the $\hat{x}$ direction, the analysis is similar: the $k_x$-axis remains a high-symmetry line and $C_{2x}$ preserves points on this line. However, since $C_{2x}$ only has two distinct eigenvalues, it can only preserve half of the possible band crossings. Thus, there can be a maximum of four Weyl points. These must all be single Weyl points: we can understand this by writing an effective two-band Hamiltonian, $H_{\rm eff} = d_i(k_x,k_y,k_z)\sigma_i$, which describes the band structure near the Weyl point. In this space, $C_{2x}$ can be represented by $i\sigma_z$ (because it squares to $-1$ and mixes bands with different $C_{2x}$ eigenvalues). Enforcing $C_{2x}$ symmetry, to lowest order, $H_{\rm eff} = (\alpha + ak_x + bk_y^2+ck_z^2)\sigma_z + (dk_y + ek_z)\sigma_x + (fk_y + gk_z)\sigma_y$. Shifting $k_x$ by $(bk_y^2+ck_z^2)/a$ and rotating $\sigma_x$ and $\sigma_y$ puts $H_{\rm eff}$ into the canonical form of a single Weyl point.

We now consider the $k_x=0$ plane, which is protected by $M_{100}$ mirror symmetry. This symmetry can protect half of the possible band crossings; the protected crossings form two line nodes in the plane. In addition, along the $k_z$ line, there is a non-topological crossing; the two-band Hamiltonian near the crossing is identical in form to Eq~(\ref{eq:nonWeyl}), which describes a band crossing in Cd$_3$As$_2$ in the same field configuration.

The same analysis applies to a magnetic field in the $\hat{y}, \hat{a}, \hat{b}, \hat{c}$ and $\hat{d}$ directions, defined in Eq~(\ref{eq:symmvecs}), because they are related by $C_{6z}$ symmetry.

We now discuss the pertubative effects of the Zeeman term: the leading order term, of order $B^2/\Delta$, contributes to the diagonal entries of the Hamiltonian, while the next order term, $B^3/\Delta^2$, mixes the $p$ orbitals; $\Delta$ is the energy gap to other $p$-orbitals outside the $k\cdot p$. This is identical to the case in Cd$_3$As$_2$.
Thus, when $B$ is along the $x$-axis, $H_Z\sim B^2/\Delta\mathbb{I}+B^3/\Delta^2\sigma_{x}$, where the matrices act on the two $p$ orbitals. These are the lowest order terms consistent with $C_{2x}$ symmetry.
Similarly, when $B$ is along the $y$-axis, $H_Z\sim B^2/\Delta\mathbb{I}+B^3/\Delta^2\sigma_{y}$.
By including $H_Z$, we have verified the claims above: when $B$ is along a high-symmetry direction, there are a maximum of four Weyl points, which are all single Weyl points, and line nodes exist in the plane perpendicular to $B$.

\subsection{Band crossings along the $k_z$ axis for arbitrary $\mathbf{B}$}

The logic at the end of the previous section regarding Cd$_3$As$_2$ applies equally well to Na$_3$Bi: in particular, symmetry restricts the Hamiltonian to be diagonal along the $k_z$ axis and thus exactly solvable.
As long as the $p$-orbitals with $|J_z| = 1/2$ are well-separated in energy from the $p$ orbitals with $|J_z|=3/2$, there will always be band crossings along the $k_z$ axis.
However, these band crossings are not always Weyl points.

\section{Magnetotransport}

\subsection{Semiclassical equations of motion}
In this section, we will derive the equations of motion governing semiclassical dynamics near a Weyl node. We work in units $\hbar=c=e=1$ throughout. We assume that we have a Fermi surface composed of $N$ \emph{disjoint} pockets surrounding Weyl points, which may be magnetic-field induced or intrinsic. The conditions for this assumption to hold in our models of field-induced Weyls are given in Eqs.~(8)--(12) in the main text. We assume that scattering between the different pockets is weak, so that to first approximation we can describe the low-energy behavior of the system with the Hamiltonian
\begin{equation}
H(\mathbf{k})=\bigoplus_i H_i(\mathbf{k})
\end{equation}
where $i=1,\dots,N$ indexes the different Weyl nodes, and
\begin{equation}
H_i(\mathbf{k})=u_{i}^\mu({k-k^{0i}})_{\mu}+(k-k^{0i})_\mu (A_i)^{\mu}_{\nu}\sigma^\nu.
\end{equation}
Here $\mu=1,2,3$ indexes the spatial direction, $k^{0i}$ is the location of the $i$-th Weyl node in the Brillouin zone, and $\sigma^\nu$ are the usual Pauli matrices. We assume that the $A_i$ are either positive definite or negative definite, so that the model describes point nodes (as opposed to line nodes). Let us introduce new coordinates 
\begin{equation}
\tilde{k}^i_\nu=(k-k^{0i})_\mu (A_i)^{\mu}_{\nu}
\end{equation}
in terms of which the linearized Hamiltonians $H_i$ may be written
\begin{equation}
H_i(\mathbf{k})=\tilde{u}_i^\mu\tilde{k}^i_\mu+\tilde{k}^i_\mu\sigma^\mu
\end{equation}
where we have introduced
\begin{equation}
\tilde{u}_i^\mu=u_i^\nu (A^{-1}_i)^\mu_\nu
\end{equation}
We assume that 
\begin{equation}
|\tilde{\mathbf{u}}_i|<1
\end{equation}
so that all our Weyl points are Type I. The spectrum of each $H_i$ is given by
\begin{equation}
\epsilon_i^{\pm}=\mathbf{u}_i\cdot\mathbf{k}^i\pm\left|\tilde{\mathbf{k}}^i\right|.
\end{equation}
where the plus sign corresponds to particle energies, and the minus sign corresponds to holes/antiparticles. The Berry curvature around each Weyl node is given by
\begin{equation}
\Omega_\mu^i=\det(A_i)\frac{ k_\mu-k^{0i}_\mu}{2|\tilde{\mathbf{k}}^i|^3},
\end{equation}
and hence the monopole charge of each node is given by
\begin{equation}
c_i=\mathrm{sgn}(\det A_i)
\end{equation}

We can now derive the semiclassical equations of motion near a Weyl node. For simplicity, we take $N=1$ for now. Our strategy, following Refs. \cite{stephanov} and \cite{xiaoreview}, will be to take the path integral representation of the propagator
\begin{equation}
K=\int\mathcal{D}\mathbf{x}\mathcal{D}\mathbf{k} e^{i\int dt (\mathbf{k}+\mathbf{A})\cdot\dot{\mathbf{x}}-\Phi-H}
\end{equation}
for a single particle (i.e. positive energy) excitation, and then perform the stationary phase approximation, with background electromagnetic vector potential $\mathbf{A}$ and scalar potential $\Phi$. For this to work, we must make sure the Weyl Hamiltonian is diagonalized at every step. Because of the path integral over momentum, this introduces the Berry connection into the classical action. After imposing the condition that classical particle trajectories have positive energy, we find that the classical action is
\begin{equation}
S_{cl}=\int dt (\mathbf{k}+\mathbf{A})\cdot\dot{\mathbf{x}}-\Phi-\epsilon^+-\mathbf{a}\cdot\dot{\mathbf{k}}
\end{equation} 
where the Berry connection $\mathbf{a}$ satisfies
\begin{equation}
\nabla\times\mathbf{a}=\mathbf{\Omega}
\end{equation}
By varying the classical action, we find for the equations of motion
\begin{align}
\dot{\mathbf{x}}&=\mathbf{v}+\dot{\mathbf{k}}\times\Omega \\
\dot{\mathbf{k}}&=\mathbf{E}+\dot{\mathbf{x}}\times\mathbf{B} \label{eoms1}
\end{align}
where $\mathbf{E}$ and $\mathbf{B}$ are the electric and magnetic field, and we have defined
\begin{equation}
v^\mu=\frac{\partial\epsilon^+}{\partial k_\mu}=u^\mu+A^{\mu}_\nu\frac{\tilde{k}_\nu}{|\tilde{\mathbf{k}}|}
\end{equation}
We can solve the equations of motion to find 
\begin{align}
\dot{\mathbf{x}}&=\frac{\mathbf{v}+\mathbf{E}\times\mathbf{\Omega}+(\mathbf{\Omega}\cdot\mathbf{v})\mathbf{B}}{1+\mathbf{B}\cdot\mathbf{\Omega}} \nonumber\\
\dot{\mathbf{k}}&=\frac{\mathbf{E}+\mathbf{v}\times\mathbf{B}+(\mathbf{E}\cdot\mathbf{B})\mathbf{\Omega}}{1+\mathbf{B}\cdot\mathbf{\Omega}} \label{eoms2}
\end{align}

\subsection{Kinetic Equation}
Using the equations of motion (\ref{eoms2}), and assuming that scattering is sufficiently weak that every node can be considered independently, we can write a Boltzmann equation for the distribution function of electrons near each node. To do so, we must first identify a (time-evolution) invariant volume element with which to define a conserved particle density\cite{XiaoPRL,Son2013PRD,xiaoreview}. Let $dV_0=d\mathbf{x}d\mathbf{k}/(2\pi)^3$ be the ``standard'' phase space volume element. By computing the Jacobian determinant for the change of variables
\begin{equation}
\mathbf{x}(t)\rightarrow\mathbf{x}(t+\delta t), \mathbf{k}(t)\rightarrow\mathbf{k}(t+\delta t)
\end{equation}
we find
\begin{equation}
dV_0(t+dt)=\left(1+\frac{\partial\dot{\mathbf{x}}}{\partial \mathbf{x}}+\frac{\partial\dot{\mathbf{k}}}{\partial\mathbf{k}}\right)dV_0(t).
\end{equation}
With the aid of the equations of motion (\ref{eoms1}) and (\ref{eoms2}), this simplifies to
\begin{equation}
dV_0(t+dt)-dV_0(t)=-\frac{dV_0(t)}{1+\mathbf{B}\cdot\mathbf{\Omega}}\frac{\partial(\mathbf{B}\cdot\mathbf{\Omega})}{\partial t},
\end{equation}
which allows us to identify
\begin{equation}
dV=dV_0(1+\mathbf{B}\cdot\mathbf{\Omega})
\end{equation}
as the time-independent volume element.

Let $n^i(\mathbf{x},\mathbf{k},t)$ be the distribution function for particles near the $i^{\rm th}$ node, defined with respect to this volume element.
% the invariant volume element $dV=d\mathbf{x}d\mathbf{k}(1+\mathbf{B}\cdot\mathbf{\omega})/(2\pi)^3$. 
Then conservation of particle number implies,
\begin{equation}
\frac{d n^i}{dt}=\frac{\partial n^i}{\partial t}+\dot{\mathbf{x}}\cdot\frac{\partial n^i}{\partial \mathbf{x}}+\dot{{\mathbf{k}}}^i\cdot\frac{\partial n^i}{\partial{\mathbf{k}}}=I^i(n^i). \label{boltzmann}
\end{equation}
 Here $I^i$ is a collision integral accounting for both inter- and intra-node scattering. The density of states $\rho^i$ near each node, defined with respect to the invariant volume, $dV$, is given by
\begin{align}
\rho^i(\epsilon)&=\int \frac{d^3k}{(2\pi)^3}\delta(\epsilon^+(\mathbf{k})-\epsilon)(1+\mathbf{B}\cdot\mathbf{\Omega}^i) \\
&=\frac{1}{4\pi^2}\left(\frac{2\epsilon^2}{|\det A_i|(1-|\tilde{\mathbf{u}}_i|^2)^2}\right. \label{rho0} \\
&\left.+\frac{c_i\tilde{\mathbf{B}}_i\cdot\tilde{\mathbf{u}}_i}{|\tilde{\mathbf{u}}|^3}(|\tilde{\mathbf{u}}_i|-\mathrm{tanh}^{-1}|\tilde{\mathbf{u}}_i|)\right),
\end{align}
where we have defined $\tilde{B}_i^\mu=(A^{-1}_i)^{\mu}_\nu B^\nu$. If we assume that the distribution functions $n^i$ depend on the momentum only through the energy (which is true in equilibrium, and is a good assumption when the intra-node scattering time is the shortest timescale in the problem; see Ref. \cite{spivak2} for more detailed discussion on this point) and that the electromagnetic field is space-independent, then we can isolate the effects of inter-node scattering by multiplying Eq. (\ref{boltzmann}) by $d^3k/(2\pi)^3\delta(\epsilon^+-\epsilon)(1+\mathbf{B}\cdot\mathbf{\Omega^i})$ on each side and integrating. Doing so yields
\begin{equation}
\frac{\partial n^i(\epsilon)}{\partial t}+\frac{1}{\rho^i(\epsilon)}\frac{\partial n^i(\epsilon)}{\partial\epsilon}\left(\frac{c_i}{4\pi^2}\mathbf{E}\cdot\mathbf{B}\right)=I^i(\epsilon), \label{nodeboltzmann}
\end{equation}
where $\rho^i_0$ is given by Eq. (\ref{rho0}), and 
\begin{equation}
I^i(\epsilon)=\frac{1}{\rho^i(\epsilon)}\int\frac{d^3\mathbf{k}}{(2\pi)^3}(1+\mathbf{B}\cdot\mathbf{\Omega})I^i(n^i)
\end{equation}
now contains only \emph{internode} scattering. For weak internode scattering, we may make the relaxation time approximation
\begin{equation}
I^i(\epsilon)=-\frac{1}{\tau}(n^i(\epsilon)-n^i_0(\epsilon)),
\end{equation}
where $n^i_0$ is the equilibrium distribution function for node $i$, and $\tau=\tau(\mathbf{B})$ is the internode scattering time, which for field-tuned Weyls may be a strong function of magnetic field. In the weak-scattering regime we are considering, it is computable in perturbation theory. We can solve Eq. (\ref{nodeboltzmann}) in this approximation to linear order in the electric field to find
\begin{equation}
n^i(\epsilon)\approx n^i_0(\epsilon)-\frac{\tau}{\rho^i(\epsilon)}\frac{\partial n^i_0(\epsilon)}{\partial\epsilon}\left(\frac{c_i}{4\pi^2}\mathbf{E}\cdot\mathbf{B}\right) \label{perturbeddist}
\end{equation}

\subsection{Anomalous conductivity of point nodes}
We will now use the distribution function Eq. (\ref{perturbeddist}) to calculate the contribution of internode scattering to the conductivity. To do so, we must first back up and derive an expression for the current density. We can do so quite generally from Eq. (\ref{boltzmann}). We know from the density of states that the number of electrons near each node is given by
\begin{equation}
N^{(i)}=\int\frac{d^3k}{(2\pi)^3}(1+\mathbf{B}\cdot\mathbf{\Omega}^i)n^i(\mathbf{k})
\end{equation}
Using this and the kinetic equation, and defining the current as
\begin{equation}
\sum_i\frac{dN^i}{dt}+\nabla\cdot\mathbf{j}=\sum_i\int\frac{d^3k}{(2\pi)^3}(1+\mathbf{B}\cdot\mathbf{\Omega}^i)I^i(\mathbf{k})
\end{equation}
we find
\begin{align}
\mathbf{j}&=\sum_i\mathbf{j}^i \\
&=\sum_i\int \frac{d^3k^i}{(2\pi)^3}\left[\mathbf{v}_i+\mathbf{E}\times\mathbf{\Omega}^i+(\mathbf{\Omega}^i\cdot\mathbf{v}_i)\mathbf{B}\right]n^i \label{currentdef}
\end{align}
Recall that $\sum_ic_i=0$ by the Nielsen-Ninomiya theorem. We can substitute Eq. (\ref{perturbeddist}) into Eq. (\ref{currentdef}) to find the internode contribution to the current at zero temperature. Using the zero-temperature result,
\begin{equation}
\frac{\partial n^i_0}{\partial \epsilon}=-\delta(\epsilon-\mu),
\end{equation}
yields
\begin{equation}
j^{i,\mu}= \frac{\tau}{16\pi^4\rho_i(\mu)}B^\mu B^\nu E_\nu.
\end{equation}
Summing over all nodes, we find for the anomalous contribution to the current
\begin{align}
    j^\mu&=\sigma^{\mu\nu}_a E_\nu \\
    \sigma^{\mu\nu}&=\frac{\tau}{16\pi^4}B^\mu B^\nu\sum_i{\rho_i(\mu)^{-1}}
\end{align}
This is the generalization of the magnetoconductance from Ref. \cite{sonspivak}, although the magnetic field dependence is now significantly more complicated, due to the magnetic field dependence of $\rho_i$. This depends on the field explicity, and also implicitly through the field dependence of $\mathbf{u}_i$ and $(A_i)^\mu_\nu$. For example, for the exactly solvable point nodes in GdPtBi with $B||(001)$ and $B||(111)$, we have $\mathbf{u}_i\propto\frac{\mathbf{B}}{\sqrt{|\mathbf{B}|}}$. Oddly enough, in one of these cases $\mathbf{B}$ is parallel to $\tilde{\mathbf{u}}_i$ as well (recall that this vector enters the density of states), though in another case the two are actually \emph{perpendicular}. Note also that, for intrinsic Weyl points, in a two-point measurement ($\mathbf{j}$ measured \emph{parallel to} $\mathbf{E}$), the measured current will depend on $\cos^2\theta$, where $\theta$ is the angle between $\mathbf{E}$ and $\mathbf{B}$, while for field-created Weyls, higher powers of $\cos\theta$ can appear.

\subsection{Simple model for field-induced Weyls}
Applying the above formalism, we consider the simplest two-band model for field induced Weyls. We take for the Hamiltonian (ignoring temporarily the orbital magnetic field)
\begin{equation}
	H=k_z^2\sigma_z+k_x\sigma_x+k_y\sigma_y-\mathbf{B}\cdot\mathbf{\sigma}
\end{equation}
At zero field, the spectrum of this Hamiltonian has a quadratic two-band touching at $\mathbf{k}=\mathbf{0}$. As the Zeeman field is increased, two Weyl nodes develop at $\mathbf{k}_{0\pm}=(B_x,B_y,\pm\sqrt{B_z})$. The linearized Hamiltonians $H_{\pm}$ around each of these points are
\begin{align}
	H_{\pm}&=\delta\mathbf{k}_{\pm}^T\mathbf{A}_{\pm}\mathbf{\sigma} \\
	\mathbf{A}_\pm&=\left(\begin{array}{ccc}
1 & 0 & 0 \\
0 & 1 & 0 \\
0 & 0 & \pm 2\sqrt{B} \end{array}\right),
\end{align}
where $\delta \mathbf{k}_\pm=\mathbf{k}-\mathbf{k}_{0\pm}$. We can now apply our analysis from the previous subsections to calculate the anomalous conductivity near each of these Weyl nodes. (This restores the orbital field, at least in the semiclassical approximation). Using Eq. (\ref{rho0}), and the fact that $\mathbf{u}=0$, we find that
\begin{equation}
	\sigma^{\mu\nu}_a=\frac{\tau(B)}{2\pi^2\mu^2}\sqrt{B_z}B^\mu B^\nu
	\label{eq:MC52}
\end{equation}
Note the additional field dependence of the current, stemming from the density of states. Furthermore, at fixed density the chemical potential will also be $B$-dependent.

Due to the simplicity of this model, we can also solve for the conductivity in the ultraquantum limit, at least for $\mathbf{B}=(0,0,B_z)$. We can then choose a gauge in which the Hamiltonian (including the orbital field) is translationally invariant along the $z$ direction. In this case, the most convenient representation for the Hamiltonian is
\begin{equation}
	H=\left(\begin{array}{cc}
	k_z^2-B_z & b^\dag \\
	b & B_z-k_z^2
	\end{array}
	\right),
\end{equation}
where $b$ is the two-dimensional Landau-level lowering operator, corresponding to the Landau levels in the $x-y$ plane. The energies of this Hamiltonian are given by
\begin{align}
    \epsilon_{\pm}&=\pm\sqrt{(k_z^2-B_z)^2+2B_zn},\;\; n=1,2,\dots \\
    \epsilon_0&=k_z^2-B_z
\end{align}
Each of these states is $B_zA_{xs}/2\pi$ - fold degenerate, coming from the degeneracy of the 2d Landau levels ($A_{xs}$ is the cross-sectional area in the $x-y$ plane). Note that around each Weyl point, the zeroth Landau level is chiral with velocity
\begin{equation}
v^\pm_z=\left.\frac{\partial\epsilon_0}{\partial k_z}\right|_{k_z=\pm\sqrt{B}}=\pm 2\sqrt{B}
\end{equation}
If we apply an external electric field $E_z$ to this system, it will shift all of the $k_z's$ as a function of time. If we assume, however, that there is a mechanism for internodal scattering with relaxation time $\tau$, then in the long-time limit, $k_z$ of the occupied states will be shifted at each node by the finite amount
\begin{equation}
	\Delta k_z=\tau E_z.
\end{equation}
This results in a depletion of right-moving states from the negative chirality node and an increase in left-moving states at the positive chirality node. Counting the states involved in this process\cite{NN}, we find a net current density
\begin{equation}
j_z=\frac{v_z B_z}{2\pi^2}\Delta k_z=\frac{\tau B_z^{3/2}}{\pi^2}E_z
\end{equation}
Note that the scaling of $B_z$ here is one power lower than in the semiclassical case (consistent with Ref~\cite{sonspivak}).

%%%%%%
%%%%%%
%%%%%%
%%%%%%
%%%%%%
%%%%%%

\subsection{Comparison to experiment}
As noted in the main text, the applicability of our theory to the existing experiments in GdPtBi is limited by the separation of the Weyl points in momentum space.
Nonetheless, in order to push our theory as far as it can go, we now show that the experimental data in Ref~\onlinecite{Hirschberger16} is better described by including higher powers of magnetic field in the magnetoconductance
%, as in the example of Eq~(\ref{eq:MC52}), 
than by the semi-classical theory of Ref~\onlinecite{sonspivak}. Below we plot magnetoconductance as a function of $B^2$ using the data from Fig 1d in Ref~\onlinecite{Hirschberger16}. 
For temperatures below 50K, the magnetoconductance has positive curvature for small $B$, indicating that it scales like a greater power than $B^2$. At higher fields, the magnetoconductance approaches the $B^2$ scaling.
Thus, the low-field data goes beyond the theory of Ref~\onlinecite{sonspivak}.

\begin{figure}[h]
\centering
\includegraphics[width=0.5\textwidth]{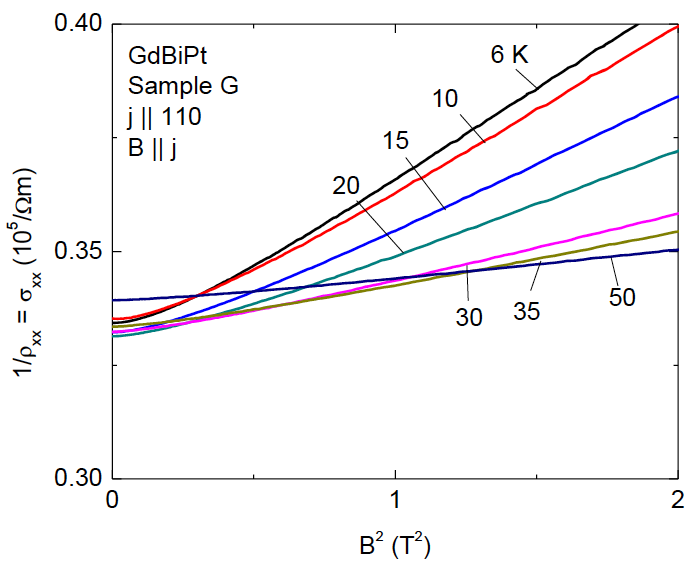}
\caption{Magnetoconductance as a function of $B^2$ for data from Fig 1d of Ref~\onlinecite{Hirschberger16}. The curvature at low fields and low temperatures shows that $\sigma_{xx}\sim B^p$, where $p>2$. At higher fields, $\sigma_{xx}\sim B^2$.}
\end{figure}

%\textcolor{blue}{(Note to us: we estimate in the main text that the Weyl points do not become resolved until $B=6$T. Hence, looking at $B\sim 1$T seems meaningless.)

\end{document}